\documentclass[a4paper, fleqn, usenatbib]{mnras}

\usepackage{newtxtext,newtxmath}
\usepackage[T1]{fontenc}
\usepackage{ae,aecompl}
\usepackage{gensymb}
\usepackage{acronym}

\usepackage{graphicx}   % Including figure files
\usepackage{amsmath}    % Advanced maths commands
\usepackage{siunitx}    %package for SI units
    \DeclareSIUnit \parsec {pc}
    \DeclareSIUnit \year {yr}
    \DeclareSIUnit \AU {AU}
    \DeclareSIUnit \erg {erg}
    \DeclareSIUnit \solarmass {M_{\odot}}
    \DeclareSIUnit \angstrom {\text {Å}}
\usepackage{dirtytalk} %package for quotation marks
\usepackage{url}
\usepackage{enumitem}
\usepackage{subcaption}
\title[Mass-metallicity relation of $z=2.2$ galaxies]{Quasar Sightline and Galaxy Evolution (QSAGE) - III. The mass-metallicity and fundamental metallicity relation of $z\approx2.2$ galaxies}

\author[H.M.O. Stephenson et al.]{H. M. O. Stephenson$^{1}$\thanks{E-mail: h.stephenson@lancaster.ac.uk (HMOS)},
J. P. Stott$^{1}$\thanks{E-mail: j.p.stott@lancaster.ac.uk (JPS)},
F. Cullen$^{2}$,
R. M. Bielby$^{3}$,
N. Amos$^{1}$,
\newauthor R. Dutta$^{4,5}$,
M. Fumagalli$^{4,6}$,
N. Tejos$^{7}$,
J. N. Burchett$^{8, 9}$,
R. A. Crain$^{10}$,
\newauthor and J. X. Prochaska$^{9}$
\\
$^{1}$Department of Physics, Lancaster University, Lancaster LA1 4YB, UK\\
$^{2}$Institute for Astronomy, University of Edinburgh, Royal Observatory, Edinburgh EH9 3HJ, UK\\
$^{3}$Department for Education, Bishopsgate House, Feethams, Darlington, DL1 5QE, UK\\
$^{4}$Dipartimento di Fisica G. Occhialini, Universit\`a degli Studi di Milano Bicocca, Piazza della Scienza 3, I-20126 Milano, Italy\\
$^{5}$INAF - Osservatorio Astronomico di Brera, via Bianchi 46, 23087 Merate (LC), Italy\\
$^{6}$INAF - Osservatorio Astronomico di Trieste, via G. B. Tiepolo 11, I-34143 Trieste, Italy\\
$^{7}$Instituto de F\'isica, Pontificia Universidad Cat\'olica de Valpara\'iso, Casilla 4059, Valpara\'iso, Chile
\\
$^{8}$Department of Astronomy and Astrophysics, UCO/Lick Observatory, University of California, 1156 High Street, Santa Cruz, CA 95064, USA\\
$^{9}$Department of Astronomy, New Mexico State University, Las Cruces, NM 88003, USA\\
$^{10}$Astrophysics Research Institute, Liverpool John Moores University, 146 Brownlow Hill, Liverpool L3 5RF, UK
}

\date{Accepted XXX. Received YYY; in original form ZZZ}

\pubyear{2023}
\begin{document}
\label{doc::firstpage}
\setlength{\abovedisplayskip}{0pt}
\setlength{\belowdisplayskip}{0pt}
\setlength{\parskip}{0pt}
\pagerange{\pageref{firstpage}--\pageref{lastpage}}
\maketitle

\begin{abstract}
We present analysis of the mass-metallicity relation (MZR) for a sample of 67 [OIII]-selected star-forming galaxies at a redshift range of $z=1.99 - 2.32$ ($z_{\text{med}} = 2.16$) using \emph{Hubble Space Telescope} Wide Field Camera 3 grism spectroscopy from the Quasar Sightline and Galaxy Evolution (QSAGE) survey. Metallicities were determined using empirical gas-phase metallicity calibrations based on the strong emission lines [OII]3727,3729, [OIII]4959,5007 and H$\beta$. Star-forming galaxies were identified, and distinguished from active-galactic nuclei, via Mass-Excitation diagrams. Using $z\sim0$ metallicity calibrations, we observe a negative offset in the $z=2.2$ MZR of $\approx -0.51$ dex in metallicity when compared to locally derived relationships, in agreement with previous literature analysis. A similar offset of $\approx -0.46$ dex in metallicity is found when using empirical metallicity calibrations that are suitable out to $z\sim5$, though our $z=2.2$ MZR, in this case, has a shallower slope. We find agreement between our MZR and those predicted from various galaxy evolution models and simulations. Additionally, we explore the extended fundamental metallicity relation (FMR) which includes an additional dependence on star formation rate (SFR). Our results consistently support the existence of the FMR, as well as revealing an offset of $0.28\pm0.04$ dex in metallicity compared to locally-derived relationships, consistent with previous studies at similar redshifts. We interpret the negative correlation with SFR at fixed mass, inferred from an FMR existing for our sample, as being caused by the efficient accretion of metal-poor gas fuelling SFR at cosmic noon.
%ZZZglobal search  for " we " and " our " " me " " I " " my "

%The abstract should briefly describe the aims, methods, and main results of the paper.
%It should be a single paragraph not more than 250 words (200 words for Letters).
%No references should appear in the abstract.
\end{abstract}


%%%%%%%%%%%%%%%%%%%%%%%%%%%%%%%%%%%%%%%%%%%%%%%%%%

%%%%%%%%%%%%%%%%% BODY OF PAPER %%%%%%%%%%%%%%%%%%%%%%%%%%%%%%%%%%%%%%%%%%%%%%%%%%%%%%%%%%%%%%%%%%%%%%%%%%%%%%%%%%%%%%%%%%%%%%%%%%%%%%%%%%%%%%%%%%

\section{Introduction}
\label{sec::intro}
%This should essentially be a mini-lit review of the state of the field, including explanations of the mass-metallicity relation, the fundamental mass metallicity relation and why it is important. Make sure to assume cosmology and stuff.

Analysis of the relationship between a galaxy's stellar mass and its gas-phase metallicity (the mass-metallicity relationship, MZR) is a key diagnostic of galaxy evolution, reflecting the complicated interplay between the formation and enrichment of stars and the inflow and outflow of gas in galaxies. The MZR shows that as stellar mass increases, galaxies have enhanced metallicities, with some studies finding the relation flattening at higher masses (e.g. \citealp{Tremonti2004,Stott2013,Zahid2014,Curti2020a}). The MZR has been studied in the literature for decades. A similar form to the MZR that showed a correlation between the magnitude of a galaxy and its gas-phase metallicity was first observed in the 1970s \citep{Lequeux1979,Garnett1987}. Magnitude was used as a substitute for stellar mass in these early works because of the difficulty in obtaining accurate mass values. Following influential works on stellar evolution (e.g. \citealp{Charlot2001,Bruzual2003}) and stellar mass determination (e.g. \citealp{Kauffmann2003a,Chabrier2003}), the MZR has now been fully established in later studies up to at least $z\sim3.5$ (e.g. \citealp{Tremonti2004,Erb2006,Mannucci2010,Stott2013,Zahid2014,Ly2016,Wuyts2016,Brown2018,Torrey2019,Sanders2021,Suzuki2021,Wang2022,Sextl2023a,Langan2023}), as well as  the stellar mass-stellar metallicity relation \citep{Cullen2019,Cullen2021,Kashino2022}. See \citet{Maiolino2019} and references therein for a comprehensive review of the MZR.

The MZR has been extended to include a dependence on star formation rate (SFR). This 3D relationship has been called the fundamental metallicity relation (FMR) and was initially discussed by \citet{Ellison2008}, before being expanded upon and fully proposed by \citet{Mannucci2010} (see also \citealp{Lara-Lopez2010}). The FMR suggested by \citet{Mannucci2010} defines a tight relationship that shows gas-phase metallicity strongly decreasing with increasing SFR for low-mass galaxies, but no SFR-dependence on metallicity for high-mass galaxies. Unlike the MZR, there is more debate surrounding the form, and even the existence of, the FMR in galaxies. \citet{Sanchez2013} studied the MZR and FMR using the Calar Alto Legacy Integral Field Area (CALIFA) survey \citep{Sanchez2012} and found that their results contradicted those of \citet{Mannucci2010} (see also \citealp{Sanchez2017}). Additionally, \citet{Barrera-Ballesteros2017} found no strong relationship between the MZR with SFR (or specific SFR, sSFR) from the Mapping Nearby Galaxies at APO (MaNGA) survey \citep{Bundy2014,Wake2016}, particularly at low SFRs. However, many other studies have found the FMR in star-forming (SF) galaxies at higher redshifts, with suggestions of a redshift evolution compared to local Sloan Digital Sky Survey (SDSS; \citealp{York2000,AdelmanMccarthy2006,Abazajian2009}) galaxies. As well as finding a strong relationship in SDSS galaxies, \citet{Mannucci2010} analysed the FMR in higher $z$ galaxies from the literature. From their results, they suggest that the FMR exists, but does not evolve, out to $z\sim2.5$ for galaxies of any stellar mass and SFR, but a large offset from the trend was found for a sample of $z\sim3.3$ galaxies from \citet{Maiolino2008} and \citet{Mannucci2009}. \citet{Mannucci2011} built upon the work done by \citet{Mannucci2010} and found a new form of the FMR that extends smoothly down to lower mass SDSS galaxies ($10^{8.3}$ \si{\solarmass}, down from a previous minimum of $\approx10^{9.2}$ \si{\solarmass}) by studying the properties of long gamma-ray burst hosts. Utilising the grism technology of the \emph{Hubble Space Telescope} (\emph{HST}) in the 3D-\emph{HST} survey \citep{Brammer2012}, \citet{Cullen2014} found the FMR in their $z\sim2.16$ galaxies but they were offset from the \citet{Mannucci2010} FMR by 0.3 dex, suggesting that an evolution in the ionization conditions between the redshifts of the samples affected the metallicity calibrations, and that the choice of metallicity indicator may affect measured values of gas-phase metallicity (see also \citealp{Teklu2020} for the latter conclusion). More recently, \citet{Li2023} used the capabilities of the \emph{James Webb Space Telescope} (\emph{JWST}) to study the MZR in dwarf galaxies between $z=2-3$ and found the FMR exists for a mass range $10^{6}-10^{10}$ \si{\solarmass} in a marginally different form than observed by \citet{Mannucci2010}, but one that is in close agreement with other studies at this redshift (see \citealp{Sanders2021}). Evidence is increasing that the MZR does have a secondary dependence with SFR (e.g. \citealp{Yates2012,Andrews2013,Stott2013,Salim2014,Salim2015,Cresci2019,Curti2020a,Baker2022,Schaefer2022}). The FMR provides a more complete picture of the processes that regulate galaxy evolution, as it accounts for the fact that there is a relationship between SFR and gas-phase metallicity at a fixed stellar mass. Further investigation of the FMR is paramount for understanding the complex mechanisms that link star formation and chemical evolution in galaxies.

The origins of the MZR and FMR are still not fully understood, and several models have been proposed to explain their existence. These include inflows and outflows of both metal-enriched \citep{Edmunds1990,Spitoni2010,Spitoni2017,Saracco2023} and metal-poor (including pristine) gas \citep{Finlator2008,Dave2010,Jimmy2015}, as well as feedback from supernovae \citep{Sakstein2011,Collacchioni2018} and Active Galactic Nuclei (AGN) \citep{Torrey2019,vanLoon2021,Yang2022}. Additionally, recent observations and simulations have suggested that the MZR also has a secondary dependence on gas mass \citep{Brown2018,DeLucia2020}, with some claiming that it is the more fundamental property linked to mass and gas-phase metallicity, and that SFR is simply a tracer for gas mass in the FMR \citep{Scholte2022}. In particular, studies have suggested that the secondary dependence is with HI mass or neutral gas fraction, as the relationship between gas-phase metallicity and SFR could be a by-product of the dependence on gas density via the Kennicutt-Schmidt relation \citep{Bothwell2013,Lagos2016,DeRossi2017}. Recently, however, \citet{Baker2023} conclude that it is in fact stellar mass that is the primary property that drives the gas-phase metallicity in galaxies, ahead of any other galaxy property including SFR, velocity dispersion and dynamical mass. Furthermore, they find that not only does gas-phase metallicity have no other significant dependence when stellar mass is included, it potentially has an anti-correlation with dynamical mass at fixed stellar mass once the primary dependence has been accounted for.

The evolution of the MZR and FMR with redshift is also an active area of research, as it can provide insights into the formation and evolution of galaxies during certain epochs of the Universe's lifetime. Several studies have investigated this and the results are mixed. Both \citet{Savaglio2005} and \citet{Erb2006} found offsets from the MZR in $z=1-2$ galaxies compared to the local $z\sim0.1$ MZR suggested by \citet{Tremonti2004}. Similar offsets from local calibrations have been repeated in many later studies (e.g. \citealp{Maiolino2008,Curti2022}). Given the proposed existence of the FMR (see above), \citet{Baker2023} suggest that any offset likely arises from the fact the MZR is tracing the SFR evolution with redshift \citep{Madau2014}. However, as previously mentioned, \citet{Mannucci2010} found no redshift evolution in the FMR out to $z\sim2.5$ but a large offset for their $z>3$ sample, whereas \citet{Cullen2014} did find an offset at $z\sim2.16$. Results from other studies have shown that there is no evolution in the FMR out to $z\sim2-2.5$ (e.g. \citealp{Henry2013,Barrera-Ballesteros2017}), and others back only a mild evolution with redshift (e.g. \citealp{Baker2022}).

Some studies suggest any perceived evolution in the FMR toward higher redshifts is a result of the choice of metallicity calibration, specifically those developed using samples of low-$z$ galaxies that may not be applicable for sources at high-$z$. Many of these calibrations use emission line diagnostics \citep{Maiolino2008,Dopita2016,Bian2018}. In their review on the use of emission lines to study galaxy evolution, \citet{Kewley2019} emphasise the need for reliable emission line diagnostics to accurately determine fundamental properties of galaxies, including their metallicities. Some of the challenges they discuss include the limited number of emission lines available to some studies, the changing ionisation structure of HII regions and the interstellar medium (ISM), and contamination from shock excitation (see \citealp{Kewley2019} for a detailed review). Some studies have sought to develop metallicity calibrations that are independent of these problems, such as \citet{Dopita2016} who developed a calibration that relies only on [NII]6585, [SII]6717,6732 and H$\alpha$ emission lines which relieves the issues caused by ISM pressure and ionisation parameter. \citet{Cullen2014} concluded that offsets from the FMR of their sample mentioned above are likely down to the empirical \citet{Maiolino2008} metallicity calibrations - derived from local SDSS galaxies - not being applicable for their $z\gtrsim2$ sample, as they do not account for changes in ionisation conditions in SF galaxies. Recently, \citet{Garg2023} used the SIMBA hydrodynamical cosmological galaxy formation simulations \citep{Dave2019} and photoionisation modelling \citep{Ferland2017,Garg2022} to analyse the redshift evolution of a number of emission line ratios used as metallicity indicators. They found that all of the emission lines they study, which are typically calibrated on $z=0$ populations of galaxies, do have some evolution with redshift, suggesting that as the galactic properties they use in their models evolve with redshift, they have a discernible impact on the line ratios. Further studies have found that metallicities of high-$z$ SF galaxies determined from locally-derived strong-line calibrations are offset from the tight sequence found in BPT diagrams of galaxies at the same redshift, suggesting some systematic offset due to the choice of calibration \citep{Steidel2014,Shapley2015,Bian2018}. 

The work in this paper aims to contribute to our understanding of the MZR and FMR by exploring them in a sample of SF galaxies at $z=1.99-2.34$ ($z_{\text{median}} = 2.16$) using near-infrared (NIR) \emph{HST} grism data from the Quasar Sightline and Galaxy Evolution (QSAGE) survey \citep{Bielby2019}. The QSAGE survey was designed to obtain hundreds of galaxy redshifts by centering \emph{HST}'s Wide-Field Camera 3 (WFC3) on 12 known quasars, all of which with pre-existing \emph{HST} UV spectra. Focussing on quasar sightline-selected fields allows the circumgalactic medium (CGM) of galaxies that are not subject to any selection effects to be studied in detail \citep{Bielby2019}. This data can be used to further develop models of galaxy feedback and fuelling mechanisms associated with the CGM (see e.g. \citealp{Bielby2017}) and investigate the effect of galaxy environments on the CGM (see e.g. \citealp{Dutta2021}). However, in this paper, we study galaxies that are generally at higher $z$ than the quasars.

This paper is arranged as follows. In Section \ref{sec::data}, the QSAGE survey is described and the sample used for this work, including how the data was reduced, is explained in detail. The methods for determining galaxy properties, namely stellar mass values (Section \ref{subsubsec::stellarmass}) and SFRs (Section \ref{subsubsec::SFRdeterm}), can be found in Section \ref{subsec::galproperties}. Section \ref{sec::mexdiagram} provides an overview of the Mass-Excitation diagram, which was used to distinguish between AGN and SF galaxy populations. Details on the metallicity calibrations used for our sample of galaxies, namely those from \citet{Maiolino2008} and \cite{Bian2018}, are found in Section \ref{subsec::maiolinocal} and \ref{subsec::biancal} respectively. The results are outlined in Section \ref{sec::results}. Discussion and conclusions are summarised in Section \ref{sec:conc}.

A standard $\Lambda$CDM cosmology model is assumed with values $\Omega_{\Lambda} = 0.7$, $\Omega_{m} = 0.3$, $H_{0} = 70$ \si{\kilo\meter\per\second\per\mega\parsec}. Any magnitudes stated are presented using the AB system. All results and models in this work assume a \citet{Chabrier2003} initial mass function (IMF) throughout and any comparison results that use a different IMF (i.e. \citealp{Salpeter1955,Kroupa2001}) are converted accordingly.

%"This trend is thought to be the result of the enrichment of the interstellar medium (ISM) by Type II supernovae (SNe II), which are the primary sources of metals in galaxies. Galaxies with higher SFRs have more SNe II, and therefore higher levels of metal enrichment in their ISMs"

%Check the above to see if it is true
%%%%%%%%%%%%%%%%%%%%%%%%%%%%%%%%%%%%%%%%%%%%%%%%%%%%%%%%%%%%%%%%%%%%%%%%%%%%%%%%%%%%%%%%%%%%%%%%

\section{Sample and Data}
\label{sec::data}

%%%%%%%%%%%%%%%%%%%%%%%%%%%%%%%%%%%%%%%%%%%%%%%%%%%%%%%%%%%%%%%%%%%%%%%%%%%%%%%%%%%%%%%%%%%%%%%%

\subsection{QSAGE}
\label{subsec::qsage}
%Here, discuss the QSAGE data in detail using data from John's 2020 paper (I think)
The data used for the analysis in this paper was taken from the QSAGE survey (\emph{HST} Cycle 24 Large Program 14594; PIs: R. Bielby, J. P. Stott; see \citealp{Bielby2019} and \citealp{Stott2020}). The QSAGE survey's main science goal was to obtain redshifts for galaxies along the lines-of-sight to quasars with pre-existing UV spectra in a redshift range $z=1.2-2.4$, with the main aim of analysing the galaxies' CGM. The survey utilised the capabilities of \emph{HST}'s WFC3 instrument, particularly the IR G141 grism, which provides useful spectra in the range of $\lambda = 10750 - 17000$ \si{\angstrom}, and also imaging from the F140W and F160W filters with spectral ranges of $\lambda = 11854 - 16129$ \si{\angstrom} and $\lambda = 13854 - 16999$ \si{\angstrom} respectively. Each target quasar field was observed with 16 grism exposures lasting approximately 1000 seconds, as well as eight $\approx 250$ \si{\second} exposures in both the F140W and F160W filters, across a total of eight \emph{HST} orbits. The G141 grism observations are the primary focus of the survey, as they provide spectroscopic data for potentially hundreds of objects simultaneously, allowing for analysis of both foreground and background galaxies around the target quasar. Imaging data, mainly from the F140W filter which has a similar spectral range to G141, is used to provide source coordinates for extracting the object spectra. As grism spectra will inevitably include contamination from nearby sources, the survey includes four separate \emph{HST} roll angles for the grism observations as well as a quantitative estimate of the contamination for each source using {\sc grizli} \citep{Brammer2009griz}. Table \ref{tab::qsageinfo} includes the location of the quasar fields in the QSAGE survey and an upper limit on the number of $z\sim2.2$ SF galaxies found in the images of each (see Section \ref{sec::mexdiagram} for explanations of \say{upper limit} and what defines an SF galaxy in this work).

% Example table. Maybe include details on the quasar fields in the sample and how many galaxies are in each of them. Discuss with John

\begin{table*}
    
    \footnotesize
    \centering
    \caption{The QSAGE sample. Field Name is the name of the central quasar of each observation but here refers to the target field for the SF galaxies. The right ascension (R.A.) and declination (Dec.) of the target quasar are derived by \citet{Stott2020}. The number of $z\sim2.2$ SF galaxies in each field (No. SF Galaxies) refers to the number of objects within the completeness limits of the sample that are deemed to satisfy the definition of star-forming via the \citet{Coil2015} Mass-Excitation diagram diagnostic (see Section \ref{sec::mexdiagram}).}
    \label{tab::qsageinfo}
    \begin{tabular}{lcccc} % 15 columns, alignment for each
        \hline
        Field Name & R.A. & Dec. & No. SF Galaxies & SF Galaxies $z$ Range\\
        \hline
QSO-J1130-1449      &   11:30:07.1  & - 14:49:27.4 & 3 & $2.29-2.32$    \\                  
LBQS-1435-0134      &   14:37:48.3  &- 01:47:10.8   & 8 & $2.00-2.23$   \\  
QSO-B1521+1009      &   15:24:24.5  & +09:58:29.1   & 11 & $2.03-2.28$  \\  
QSO-B1634+7037      & 16:34:29.0 &  +70:31:32.4 & 3 & $2.16-2.17$   \\      
PKS-0232-04 &   02:35:07.3  &   -04:02:05.3 & 5 & $2.18-2.32$   \\  
QSO-B1630+3744  &   16:32:01.1  &   +37:37:50.0 & 4 & $2.06-2.21$   \\              
PG0117+213      &   01:20:17.3  &   +21:33:46.2 & 3 & $2.12-2.13$   \\   
QSO-B0810+2554  &   08:13:31.3  &   +25:45:03.1 & 5 & $2.15-2.30$   \\
HE0515-4414 &  05:17:07.6   &   -44:10:55.6 & 5 & $1.99-2.30$       \\      
2QSO-B0747+4259     & 07:50:54.6    &   +42:52:19.3 &   3 & $2.02-2.26$ \\%
QSO-J1019+2745  &  10:19:56.6   &   +27:44:01.7 &   12 & $2.02-2.24$    \\
QSO-B1122-168       & 11:24:42.9    &   -17:05:17.4 &   5 & $2.03-2.16$ \\%

        \hline
    \end{tabular}
\end{table*}

\subsection{Sample Selection}
\label{subsec::sampleselec}

In order to select only those $z\sim2.2$ objects relevant to the analysis in this work, the following criteria must be met:

\begin{enumerate}[leftmargin=*]
  \item The objects are primarily selected based on their [OIII]5007 emission. [OIII]5007 emission flux was chosen as the primary selection criterion because this line is much stronger compared to other lines at the target redshift of this paper (namely [OII]3727 and H$\beta$). Therefore the objects must have a clean spectrum with [OIII]5007 coverage. Given the need for multiple emission lines to determine gas-phase metallicity and distinguish ionisation processes, objects must also have a spectroscopic [OIII]5007 redshift that puts [OII]3727 and H$\beta$ within the wavelength range of the G141 grism. The definition of \say{clean spectrum} is explained by \citet{Stott2020} as being galaxies with a \say{quality flag} of 3 or 4. As defined in \citet{Stott2020}, \say{3} is a good quality spectrum with at least one spectral line having S/N > 3, and \say{4} is spectrum with lines that have S/N > 10. This cut was necessary in order to make sure that any emission line of an object could be reliably used when calibrating its gas-phase metallicity and used in ionisation mechanism diagnostics (see Section \ref{sec::mexdiagram}).
  \item Additionally, objects must have an [OIII]5007 emission line flux with a signal-to-noise (S/N) $\geq 3$. S/N here is defined as the [OIII]5007 flux value over the flux error.
\end{enumerate}

\begin{figure*}
\centering

\begin{subfigure}{0.45\linewidth}
\includegraphics[width=\linewidth, trim=0 0 50 0]{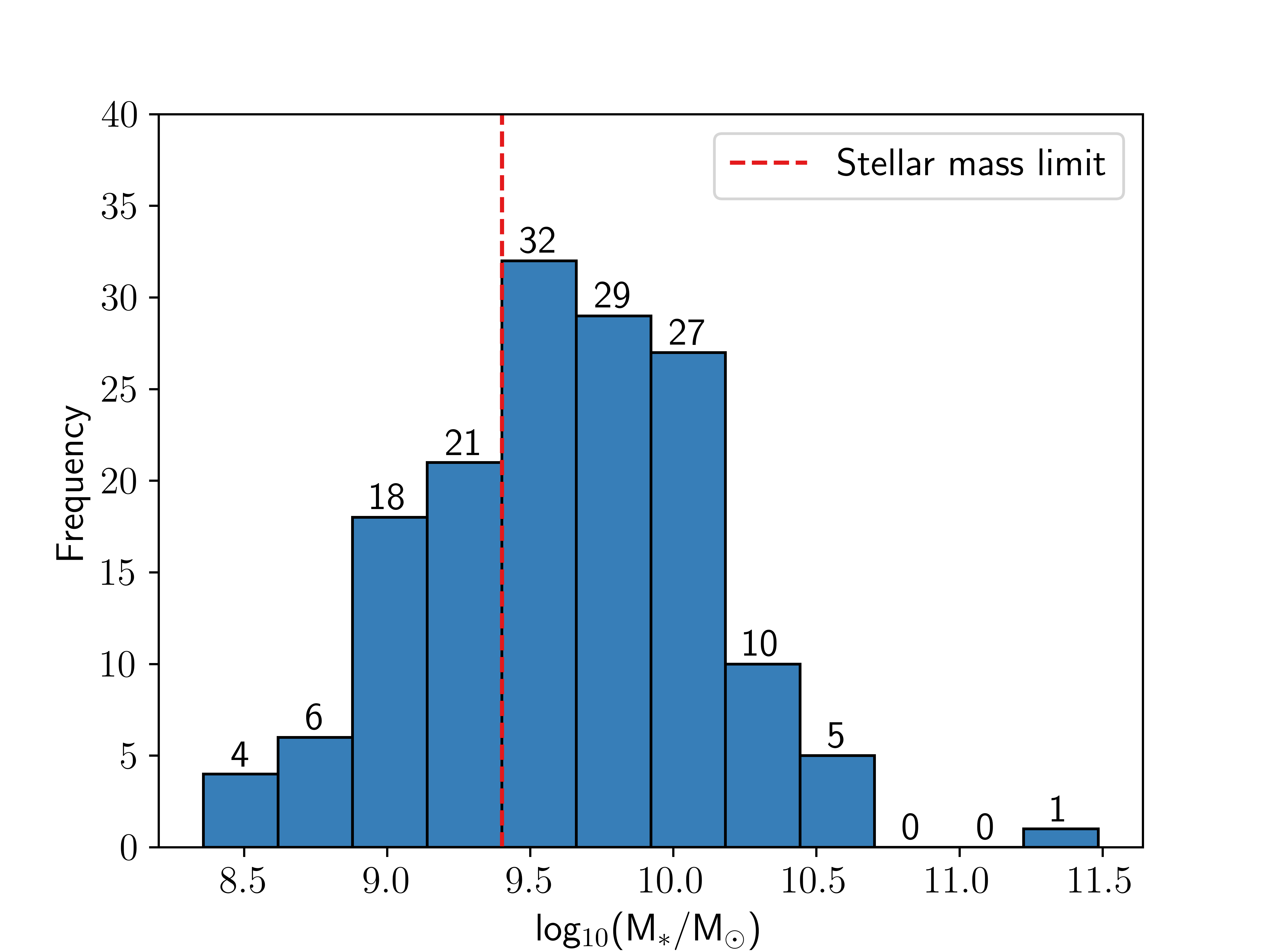}
\caption{}
\label{subfig::totalmasshist}
\end{subfigure}
\hfill
\begin{subfigure}{0.45\linewidth}
\includegraphics[width=\linewidth, trim=50 0 0 0]{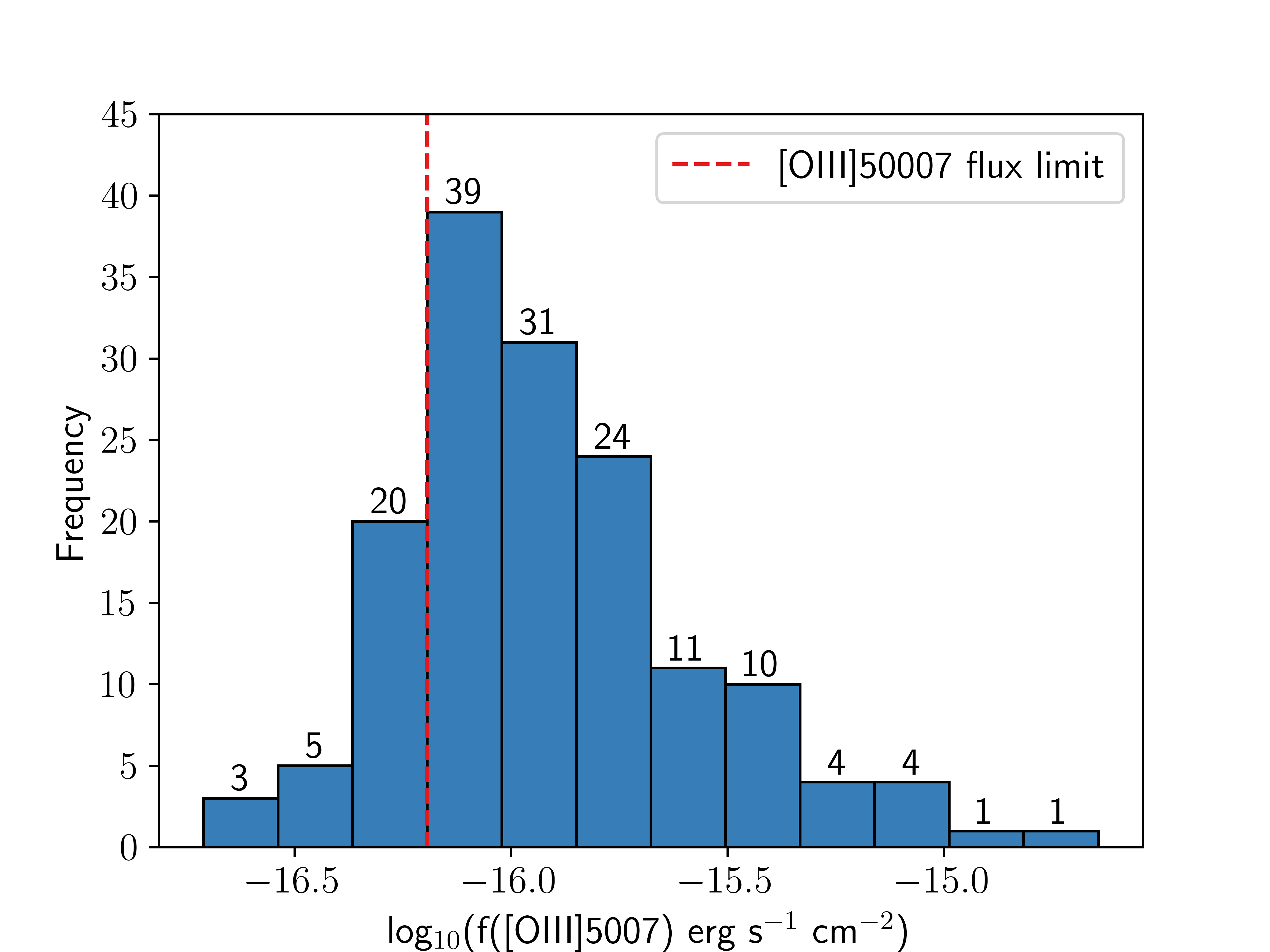}
\caption{}
\label{subfig::totaloiiifluxhist}
\end{subfigure}

\caption{The stellar mass and [OIII]5007 flux distributions of the objects that met the selection criteria in Section \ref{subsec::sampleselec}. \emph{Left} - The stellar mass distribution of the sample, where the red dashed line indicates the approximate mass-complete limit of the sample $\sim10^{9.4}$ \si{\solarmass}. \emph{Right} - The [OIII]5007 flux distribution of the sample, where the red dashed line indicates the flux-complete limit of the sample $\sim 10^{-16.2} \si{\erg\per\second\per\centi\meter\squared}$. The numbers above each bin indicate the number of objects that occupy that bin.}
\label{fig::massandfluxdistribution}
\end{figure*}

These selection criteria cut the original sample of 1953 objects down to 153. It was clear, however, that this new sample was incomplete for lower valued bins in both [OIII]5007 flux and stellar mass, so an additional completion limit was then applied to this sample of 153 objects. In this case, that was any object with $\log_{10}(f([\text{OIII}]5007$) \si{\erg\per\second\per\centi\meter\squared}) $\gtrsim -16.2$ and $\log_{10}(\text{M}_{*}/\text{M}_{\odot}) \gtrsim 9.4$, which are the lower values of the most occupied bin in each property (see Figure \ref{fig::massandfluxdistribution}). The [OIII]5007 flux and stellar mass distribution of this sample of selected objects, as well as the completeness limit, can be seen in Figure \ref{fig::massandfluxdistribution}. Statistical analysis found that both the stellar mass and [OIII]5007 flux cuts represent a $\sim95\%$ completeness limit of the original sample. After applying these, there were 92 objects in total for analysis within a redshift range of $1.99 \lesssim z \lesssim 2.32$.

%%%%%%%%%%%%%%%%%%%%%%%%%%%%%%%%%%%%%%%%%%%%%%%%%%%%%%%%%%%%%%%%%%%%%%%%%%%%%%%%%%%%%%%%%%%%%%%%

\subsection{Galaxy Properties}
\label{subsec::galproperties}

%%%%%%%%%%%%%%%%%%%%%%%%%%%%%%%%%%%%%%%%%%%%%%%%%%%%%%%%%%%%%%%%%%%%%%%%%%%%%%%%%%%%%%%%%%%%%%%%

\subsubsection{Galaxy Stellar Mass Calibration}
\label{subsubsec::stellarmass}

The QSAGE masses are inferred using an approximation based on their apparent magnitude in the IR. For this study, the calibrations from \citet{Stott2020} are used. They found linear relationships between stellar mass and F160W magnitude at different redshifts based on galaxies from the Cosmic Assembly Near-infrared Deep Extragalactic Legacy Survey (CANDELS; \citealp{Barro2019}). They derive the relationships in individual redshift bins of $\Delta z = 0.1$. The parameters they use can be found in Table A1 of their paper. Of the 92 objects in our complete sample, two do not have F160W coverage (both found in the QSO-J1130-1449 field). The F160W magnitudes for these two objects were estimated based on a linear fit between the F140W and F160W magnitudes of all the $z\sim2.2$ objects with a spectra flagged as \say{good}.

As described above, \citet{Stott2020} demonstrate a strong correlation between F160W and full SED-fit stellar mass for SF galaxies in the CANDELS survey. In the absence of deep homogeneous data across all 12 of the QSAGE fields, we chose to use this F160W magnitude calibration (corresponding to rest frame V-Band). This is redward of the 4000 \si{\angstrom} break and is therefore less affected by ongoing SF than shorter wavelengths. The primary samples we use to analyse the MZR and FMR are all SF galaxies, which means they will all have similar mass-to-light ratios. We fully account for the scatter in the \citet{Stott2020} relationship in our stellar mass errors (typical stellar mass error can be seen in the lower right corner of Figure \ref{fig::mexdiagram}).

%%%%%%%%%%%%%%%%%%%%%%%%%%%%%%%%%%%%%%%%%%%%%%%%%%%%%%%%%%%%%%%%%%%%%%%%%%%%%%%%%%%%%%%%%%%%%%%%
\subsubsection{Determining SFRs}
\label{subsubsec::SFRdeterm}

\begin{figure}{}
        \centering
        \includegraphics[width=\columnwidth, trim=0 0 0 0, clip=true]{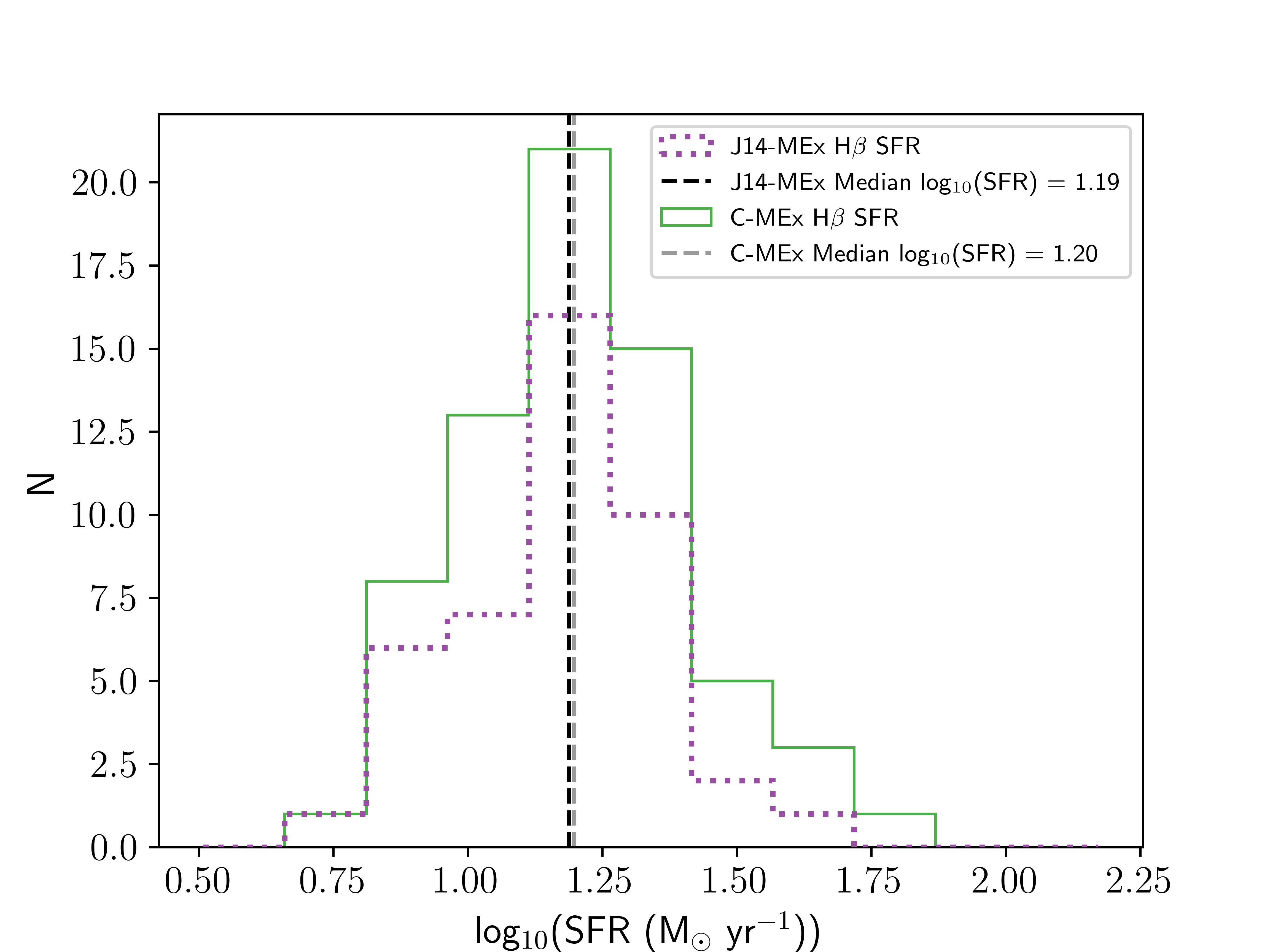}
    \caption[]{The SFR distribution of our complete J14-MEx (purple dotted histogram) and C-MEx (solid green histogram) SF galaxy samples. The dashed black line indicates the median SFR of our J14-MEx sample, and the grey dashed line shows the median SFR of the C-MEx sample. The SFR distributions were derived using the H$\beta$ emission line flux \citep{Kennicutt1998}.}
        \label{fig::sfrdistribution}
\end{figure}{}

The FMR describes the dependence of gas-phase metallicity on both stellar mass and SFR \citep{Ellison2008,Mannucci2010}. In order to analyse the FMR using the samples in this work, the SFRs of each galaxy are needed. The SFRs were calculated using the \citet{Kennicutt1998} calibrations, making sure to convert from a \citet{Salpeter1955} to a \citet{Chabrier2003} IMF by dividing the resulting SFR by a conversion factor of 1.59 \citep{Madau2014}. SFR was calculated using the H$\beta$ line luminosity, and then using the common conversion factor of H$\alpha$/H$\beta$ = 2.86 \citep{Gaskell1984,Osterbrock2006} to estimate the SFR. The theoretical value of the Balmer decrement H$\alpha$/H$\beta$ = 2.86 is used as the spectra in our sample do not have H$\alpha$ coverage at this redshift. The emission line luminosity was extinction corrected assuming $A_{V} = 1$, which, based on the properties of our sample, was most appropriate following the extinction relationships of \citet{Sobral2012}. In order to correct for underlying stellar absorption, H$\beta$ flux values were boosted by 3\%, in line with other studies into the FMR \citep{Henry2013,Cullen2021,Sanders2021,Curti2022}. These SFRs, calculated using the H$\alpha$ calibration from \citet{Kennicutt1998}, will be referred to as H$\beta$ SFR hereafter. The SFR distribution of our sample can be found in Figure \ref{fig::sfrdistribution}.

Figure \ref{fig::sfrmainseq} shows the correlation between SFR and stellar mass for individual SF galaxies in our sample. The grey dashed line shows the \say{UV+IR/IR SFRs} best-fit main sequence evolution for galaxies out to $z\approx6$ as found by \citet{Speagle2014}, with the grey shaded region indicating the scatter based on the errors on their relationship. From Figure \ref{fig::sfrmainseq}, our sample is consistent with the main sequence from \citet{Speagle2014}.

\begin{figure}{}
        \centering
        \includegraphics[width=\columnwidth, trim=0 0 0 0, clip=true]{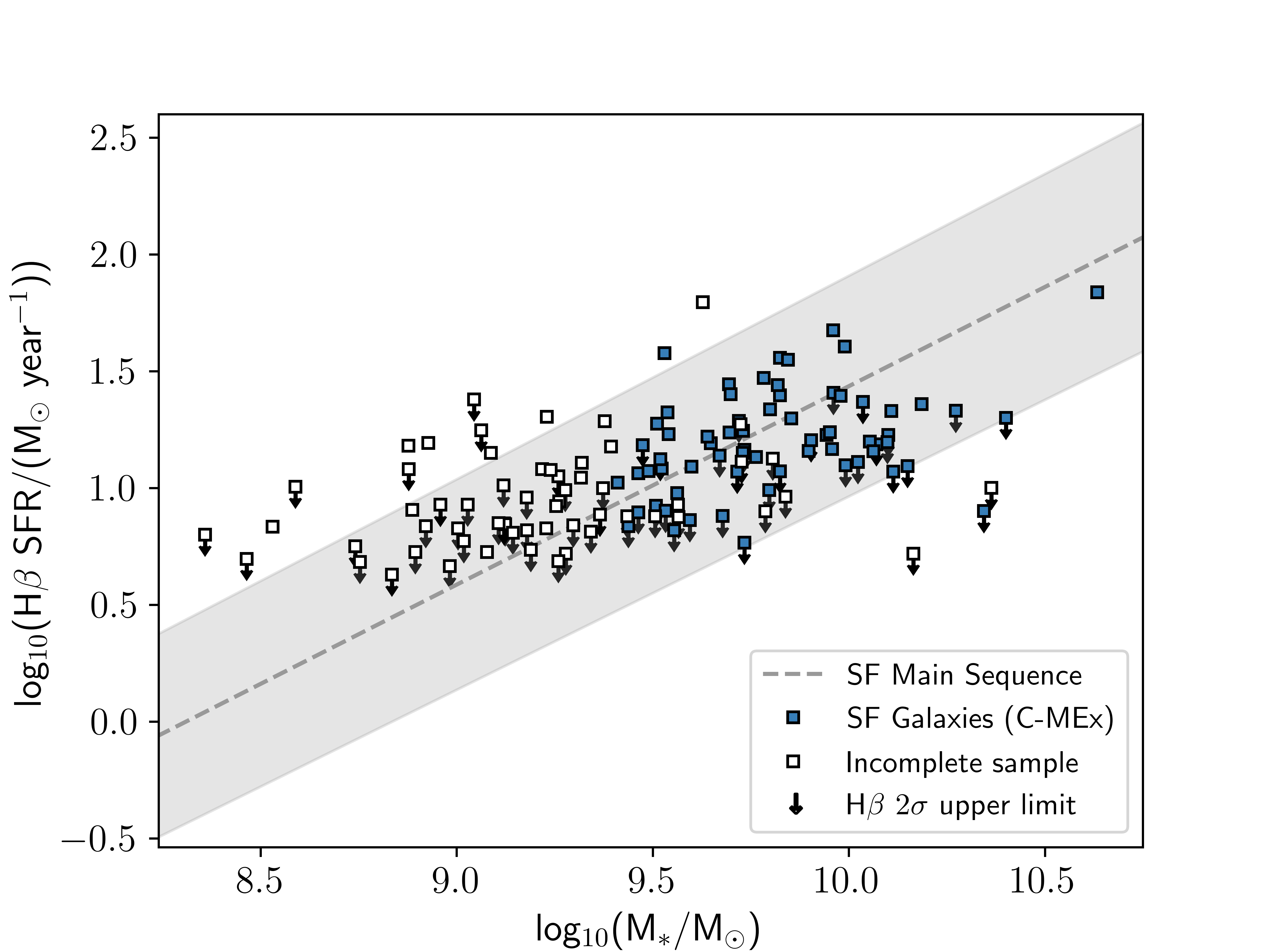}
    \caption[]{The relationship between stellar mass and SFR for the SF galaxies in our C-MEx sample. The grey dashed line shows the ``UV+IR/IR'' star-forming main sequence as described by \citet{Speagle2014}, with the grey shaded region indicating the scatter around the fit based on the errors of their equation. The blue squares show the individual galaxies in the mass and flux complete sample of our work, with white squares showing SF galaxies in the incomplete sample.}
\label{fig::sfrmainseq}
        \label{fig::sfrmainseq}
\end{figure}{}

%%%%%%%%%%%%%%%%%%%%%%%%%%%%%%%%%%%%%%%%%%%%%%%%%%%%%%%%%%%%%%%%%%%%%%%%%%%%%%%%%%%%%%%%%%%%%%%%

\section{Mass-Excitation Diagram}
\label{sec::mexdiagram}

The selection criteria listed in Section \ref{subsec::sampleselec} does not distinguish between object spectra that belong to an AGN or those that belong to SF galaxies - the primary target of the analysis in this paper. Therefore, diagnostics to differentiate the two populations must be employed so as to not contaminate the sample. Traditionally, researchers have used \say{BPT} diagrams \citep{Baldwin1981} to distinguish between AGN and SF galaxies. These diagrams use different emission line ratios to distinguish the ionisation mechanism of gas in the ISM. However, as previously discussed, the emission lines used in our analysis are [OIII]4959,5007, [OII]3727 and H$\beta$ and therefore no form of BPT diagram can be used to categorise these populations.

\citet{Juneau2011} discovered that comparing the line ratio [OIII]5007/H$\beta$ (hereafter O3) against the total stellar mass of the object is successful at distinguishing between ionisation from AGN and that from active SF. They call this the Mass-Excitation (MEx) diagram which finds that the BPT-SF and BPT-AGN classifications are well separated in this plane, including finding that BPT-composites (populations that have some combination of SF and AGN contributions) are reasonably well-defined in a central region known as \say{MEx-intermediates}. The distinct classes in the MEx diagram come from two empirical curves that maximise this separation. However, \citet{Juneau2011} notes that this diagnostic was only applied at intermediate redshifts ($0.3 < z < 1$) and may not be accurate for higher redshift samples. Studies have found that the \citet{Juneau2011} MEx diagram (hereafter J11-MEx) does hold up to $z\lesssim1.6$ (e.g. \citealp{Juneau2013,Trump2013}), but it generally fails at accurately distinguishing SF galaxies from AGN at $z\sim2$ (e.g. \citealp{Newman2013,Henry2013}).

While \citet{Juneau2011} used a S/N criterion that applied to the individual emission lines (requiring S/N > 3 on all lines), \citet{Juneau2014} developed a new MEx diagram (hereafter J14-MEx) which instead requires all \emph{line ratios} to have a S/N > $3/\sqrt{2} \approx 2.12$ (equivalent to an average $3\sigma$ detection at the lower limit for the individual lines in the ratio), resulting in a 20\% larger sample. Additionally, \citet{Juneau2014} makes use of SDSS DR7 \citep{Abazajian2009} for their low redshift sample, rather than SDSS DR4 \citep{AdelmanMccarthy2006} used for the original J11-MEx. However, the most important update to the J14-MEx is its application to higher redshift samples ($z\gtrsim2$) by taking into account the evolution of galaxy populations, in particular the fading of the luminosity function of emission-line galaxies toward lower redshifts (e.g. \citealp{Colbert2013,Khostovan2015,Hayashi2020}). They find that as the cut-off line luminosity is raised, the split between AGN and SF galaxies occurs at higher masses and so they employ a mass-offset as a function of the threshold line luminosity of a sample following the form
\begin{equation*}
\Delta\log_{10}(M_{*}/M_{\odot}) = a_{0} + a_{1} \times \tan^{-1}((\log_{10}(L_{\text{threshold}})-a_{2}) \times a_{3}),
\end{equation*}

\noindent with coefficients \{0.28988, 0.28256, 40.479, 0.82960\}. Using the line flux limit of the QSAGE survey ($f > 2\times10^{-17}$ \si{\erg\per\second\per\centi\meter\squared}; \citealp{Bielby2019}) and the median redshift of our complete sample ($z\sim2.16$), the mass offset on the J14-MEx for this sample is $\Delta\log_{10}(M_{*}/M_{\odot})\approx0.25$ which agrees with offsets calculated by other studies with similar flux limits and redshifts (e.g. \citealp{Coil2015}).

\begin{figure}
        \centering
        \includegraphics[width=\columnwidth, trim=0 0 0 0, clip=true]{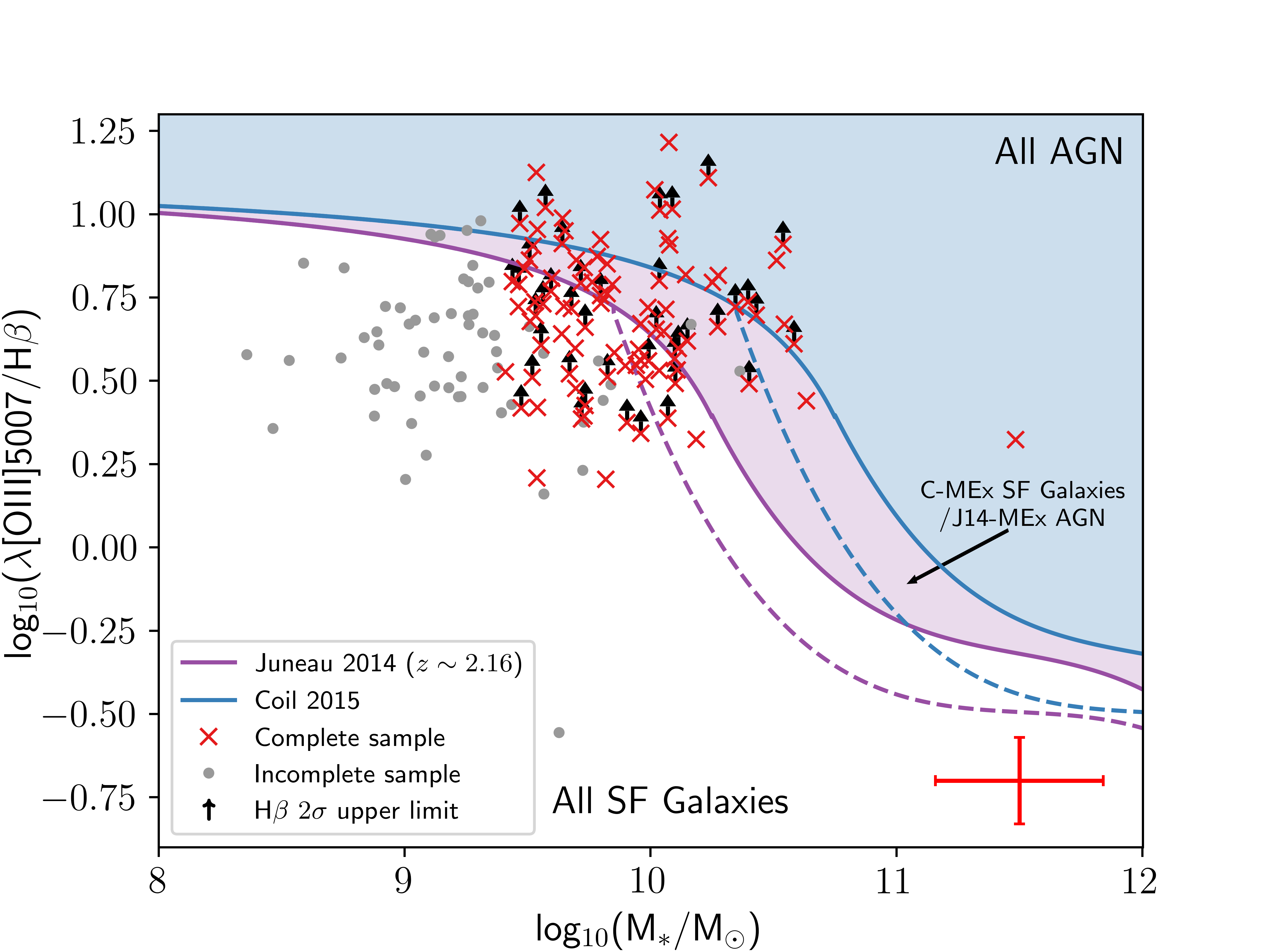}
    \caption[]{The Mass-Excitation diagram used to distinguish between ionisation from AGN and that from SF. The red crosses are objects that are within the completeness limits of both stellar mass and [OIII]5007 flux, and grey points are those that lie outside of one or both. Black arrows indicate those objects in our complete sample that have a H$\beta$ upper limit (lower limit on O3). Both of the curves show the MEx prescriptions used to distinguish ionisation processes for our sample, namely the mass-shifted (0.25 dex) \citet{Juneau2014} curve that is more applicable to higher redshifts (purple), and the \citet{Coil2015} curve (blue) that shifted the \cite{Juneau2014} curve further (0.75 dex). The area in-between the dashed and solid curves indicate a population of \say{MEx-intermediates} where there is some combination of SF and AGN activity (analogous to BPT-composites which are reasonably well-defined in this region). For our samples, we considered these MEx-intermediates to be SF galaxies. Objects occupying the unshaded region below the solid \citet{Juneau2014} MEx curve are always considered SF galaxies, and objects occupying the blue shaded region above the solid \citet{Coil2015} MEx curve are always considered AGN. In the purple shaded region between the two curves, objects are considered to be SF galaxies for the C-MEx diagnostic (because they lie below the solid blue curve), but are considered to be AGN for the shifted J14-MEx (because they lie above the solid purple line). The red error bars in the bottom right corner show the typical errors of the objects.}
        \label{fig::mexdiagram}
\end{figure}

This mass shift of 0.25 dex may be insufficient to clearly separate high-$z$ AGN and SF galaxies, despite the prescriptions in \citet{Juneau2014}. \citet{Coil2015} also found a shift of 0.25 dex to the MEx diagram for their sample but found that this shift still leaves many known SF galaxies in the AGN region (see their Figure 5). Instead, \citet{Coil2015} found that a shift of $\Delta\log_{10}(M_{*}/M_{\odot}) = 0.75$ is needed and results in all ten of their X-ray and IR-selected AGN being consistent with occupying the AGN region of the J14-MEx diagram. They note that other studies with samples at a similar redshift (e.g. \citealp{Newman2013,Price2014}) found that a similarly substantial shift of 0.75 dex is needed for the original J11-MEx, suggesting the prescriptions in \citet{Juneau2014} are generally insufficient for samples at $z\gtrsim2$. For completeness, the analysis of the MZR and FMR in this study will look at samples of SF galaxies determined using both a 0.25 dex and 0.75 dex shift to the \citet{Juneau2014} prescriptions (43 and 67 SF galaxies respectively), with the primary focus being on the \citet{Coil2015} 0.75 dex shift (hereafter the C-MEx sample). Figure \ref{fig::mexdiagram} shows the MEx diagram for the sample of objects in this paper with both the shifted \citet{Juneau2014} and \citet{Coil2015} curves discussed in this section highlighted. For the purposes of this paper, an object was considered a SF galaxy if it was below the upper solid line of the empirical curves, including those that would otherwise be considered as MEx-intermediates. We used a 2$\sigma$ upper limit on the value of H$\beta$ here (lower limit on O3) where the S/N < 2, which means the number of SF galaxies determined via this method is also an upper limit.

The stellar mass range of our J14-MEx sample of 43 SF galaxies is M$_{*}$ = 10$^{9.41 - 10.19}$ \si{\solarmass} (median stellar mass M$_{*}$ = 10$^{9.70}$ \si{\solarmass}), and the range for our C-MEx sample of 67 SF galaxies is M$_{*}$ = 10$^{9.41 - 10.63}$ \si{\solarmass} (median stellar mass M$_{*}$ = 10$^{9.80}$ \si{\solarmass}). The SFR range of our J14-MEx sample is 5.8-37.7 \si{\solarmass\per\year} (median SFR of 15.4 \si{\solarmass\per\year}), and the range for our C-MEx sample is 5.8-68.9 \si{\solarmass\per\year} (median SFR of 15.7 \si{\solarmass\per\year}).

%%%%%%%%%%%%%%%%%%%%%%%%%%%%%%%%%%%%%%%%%%%%%%%%%%%%%%%%%%%%%%%%%%%%%%%%%%%%%%%%%%%%%%%%%%%%%%%%

\section{Determining Gas-phase Metallicity}
\label{sec::metaldetermin}

Gas-phase metallicity at high redshifts ($z > 1$) could only be properly determined using strong-line metallicity diagnostics, as measurements of the electron temperature ($T_{e}$, often used to directly measure metallicities in local galaxies) become difficult due to the faintness of temperature sensitive emission lines \citep{Sanders2020}. However, the resolution and spectroscopic capabilities of \emph{JWST} is now making this possible (e.g. \citealp{Curti2022,Sanders2023b}), meaning studies using it will not necessarily have to rely on strong-line calibrations to measure metallicity of these high-$z$ galaxies. Strong-line diagnostics follow the relationship between optical emission line ratios and the abundance of heavy elements in a galaxy. It is important to note that analysis into the MZR (or other metallicity relations) is strongly dependant on the choice of calibration as it can result in significantly different curve shapes and intercepts (see \citealp{Kewley2008} for a detailed review). It is therefore crucial to use the same metallicity calibration as those comparison studies. Here, two key calibrations are adopted for analysing the MZR and FMR. They are the calibration curves from \citet{Maiolino2008}, who developed strong-line relations from a combination of metallicities measured using the $T_{e}$ method in \citet{Nagao2006} and photoionization models on SDSS DR4 galaxies \citet{AdelmanMccarthy2006}, and those from \citet{Bian2018} who developed empirical calibrations from local galaxies with properties analogous to galaxies at $z\sim2$.

%%%%%%%%%%%%%%%%%%%%%%%%%%

\subsection{Maiolino et al. (2008) Calibrations}
\label{subsec::maiolinocal}

The \citet{Maiolino2008} calibrations were needed in order to compare the MZR and FMR determined for our sample to those by \citet{Mannucci2010} and \citet{Cullen2014}. As described above, \citet{Maiolino2008} used a combination of directly measured metallicities via the $T_{e}$ method and photoionization models. However, it is well documented that the $T_{e}$ method fails at high metallicities (reliable for $12 + \log_{10}(\text{O/H})\lesssim8.3$, see \citealp{Stasinska2005,Bresolin2007}) as a result of temperature fluctuations. Photoionisation models are also prone to systematic effects as well as high uncertainties and often do not reliably reproduce expected trends at low metallicities (e.g. \citealp{Dopita2006,DorsJr2011}). Therefore, \citet{Maiolino2008} combines a low-metallicity sample from \citet{Nagao2006} that used direct gas-phase metallicity measurements via $T_{e}$, and a high-metallicity sample using galaxies from SDSS DR4 \citep{AdelmanMccarthy2006} with the photoionization models described in \citet{Kewley2002}. They then fitted polynomial curves to various strong-line ratio relationships against gas-phase metallicity after binning the galaxies in $\sim 0.1$ dex metallicity bins. The key relationships for our sample were those using $R_{23} = ([\text{OIII}]5007 + [\text{OIII}]4959 + [\text{OII}]3727)/\text{H}\beta$, $[\text{OIII}]5007/[\text{OII}]3727$ (herein O32), $[\text{OII}]3727/\text{H}\beta$ (herein O2) and O3. For the [OIII]4959,5007 doublet, the QSAGE team fitted a triple Gaussian to determine the line flux. However, due to the grism's spectral resolution constraints, the peaks are only marginally resolved in most spectra and so a fixed flux ratio of $[\text{OIII}]5007/[\text{OIII}]4959 = 2.98$ was assumed \citep{Storey2000}. This fit also included H$\beta$ ($\lambda$ = 4861 \si{\angstrom}) since its close proximity may have led to some blending \citep{Stott2020}. The general form of the polynomials fitted by \citet{Maiolino2008} is

\begin{equation}
\label{eq::maiolinofunc}
    \log_{10}(R) = c_{0} + c_{1}x + c_2x^2 + c_3x^3 + c_4x^4,
\end{equation}
where $R$ is the strong-line ratio, $x = 12 + \log_{10}(\text{O/H}) - 8.69$ and $c_{n}$ are coefficients that depend on the chosen ratio (see Table 4 in \citealp{Maiolino2008}).

We calibrated metallicities by following a $\chi^{2}$ minimisation approach adopted by \citet{Sanders2021} (see also \citealp{Cullen2021,Curti2023}). This method involves minimising the $\chi^{2}$ of multiple line ratios simultaneously using the formula
\begin{equation}
\label{eq::chisquared}
    \chi^{2}(x) = \sum_{i}\frac{(R_{\text{obs,}i} - R_{\text{cal,}i}(x))^{2}}{(\sigma_{\text{obs,}i}^{2} + \sigma_{\text{cal,}i}^{2})},
\end{equation}

\noindent where the sum over $i$ represents the set of the line ratios used for the gas-phase metallicity determination, $x = 12 + \log_{10}(\text{O/H})$, $R_{\text{obs,}i}$ is the logarithm of the $i$th observed line ratio, $R_{\text{cal,}i}(x)$ is the predicted logarithmic value of $R_{i}$ at $x$ from the \citet{Maiolino2008} calibrations, $\sigma_{\text{obs,}i}$ is the uncertainty in the $i$th observed line ratio, and $\sigma_{\text{cal,}i}$ is the uncertainty of the $i$th line ratio at a fixed $x$ of the \citet{Maiolino2008} calibrations. For $\sigma_{\text{cal,}i}$, we used the values from Table 2 of \citet{Sanders2021} which are the average values for multiple calibrations (including \citealp{Maiolino2008}), but they note that all of them have similar scatter for each of the line ratios used. Metallicities were calculated by selecting the value that minimised Equation \ref{eq::chisquared}. To determine the uncertainty of the gas-phase metallicity, we followed the method of \citet{Cullen2021}; the observed line ratios were perturbed by their uncertainty values using a Gaussian distribution and the metallicity that minimised Equation \ref{eq::chisquared} was recalculated 500 times. The 1$\sigma$ uncertainty on the calculated metallicity was then derived from the 68th percentile width of the resulting distribution of perturbed metallicities.

Since [OII]3727 and H$\beta$ may be undetected in the QSAGE sample (S/N < 2), the strong-line ratios used to calibrate the gas-phase metallicity must be carefully chosen. The reader should be reminded that [OIII]5007 is detected to S/N > 3 in all cases (see Section \ref{subsec::sampleselec}). The selection process for the ratio(s) used in Equation \ref{eq::chisquared} is as follows:

\begin{enumerate}[label=\textbf{\roman*}), leftmargin=*]
  \item If [OII]3727 and H$\beta$ are both detected with S/N $\geq 2$, then all of R$_{23}$, O3, O2 and O32 are used in Equation \ref{eq::chisquared} (applies to 28/67 C-MEx SF galaxies).
  \item If [OII]3727 is poorly detected (S/N < 2), but H$\beta$ is well detected (S/N $\geq 2$) then only O3 is used (9/67 C-MEx SF galaxies).
  \item If H$\beta$ is poorly detected (S/N < 2), but [OII]3727 is well detected (S/N $\geq 2$) then only O32 is used (26/67 C-MEx SF galaxies).
  \item In the event that both H$\beta$ \emph{and} [OII]3727 are poorly detected (S/N $<2$), then the $\chi^{2}$ minimisation method is not followed. Instead, we solve Equation \ref{eq::maiolinofunc} using O32 with an upper limit on [OII]3727 (4/67 C-MEx SF galaxies). The upper limit was applied as follows: if [OII]3727 flux was poorly detected (i.e. $0 < \text{S/N} < 2$), then the flux was corrected to double the value of the uncertainty on the measurement i.e. a 2$\sigma$ upper limit. If no [OII]3727 was detected, then the [OII]3727 flux value was set to be double the value of the corresponding [OIII]5007 flux error for that object; if the [OII]3727 upper limit results in an O32 value that is on the curve described by Equation \ref{eq::maiolinofunc}, then this metallicity was selected (upper limit on metallicity). Should an object have an O32 value that is above the curve, then the metallicity corresponding to the maximum strong-line ratio on the curve would have been selected (and these noted with extreme caution as having a lower metallicity limit), but there were no such cases in this sample.
\end{enumerate}

Since some of the curves defined by Equation \ref{eq::maiolinofunc} have multiple solutions, O32 was used to discriminate between them since the shape of the O32 curve in the metallicity range $7 \lesssim 12 + \log_{10}(\text{O/H}) \lesssim 9.5$ has a single solution.

%%%%%%%%%%%%%%%%%%%%%%%%%%%%%%%%%%%%%%%%%%%%%%%%%%%%%%%%%%%%%%%%%%%%%%%%%%%%%%%%%%%%%%%%%%%%%%%%

\subsection{Bian et al. (2018) Calibrations}
\label{subsec::biancal}

To compare our MZR and FMR results to higher redshift studies, we need to adopt the higher redshift metallicity calibrations these studies used. We primarily look at the FMR of \citet{Li2023} who recently analysed the gas-phase metallicity of galaxies at $z=2-3$ using \emph{JWST} (also to their MZR, see Section \ref{subsec::mzrelation}). In order to compare to \citet{Li2023}, the metallicity calibrations from \citet{Bian2018} must be used. \citet{Bian2018} selected a sample of local galaxies from the MPA-JHU value-added catalogue of SDSS DR7\footnote{https://wwwmpa.mpa-garching.mpg.de/SDSS/DR7/} \citep{Abazajian2009} that are analogous to $z\sim2$ SF galaxies on the O3 versus [NII]6584/H$\alpha$ BPT diagram, making these calibrations more applicable to high redshift samples. \citet{Bian2018} used definitions of SF on the BPT diagram as defined in \citet{Kewley2013} for the local reference sample and \citet{Steidel2014} for the high-$z$ analogue sample (using \citealp{Kewley2006} criterion to remove AGN and shock contamination to the emission line flux). The \citet{Bian2018} calibrations used for our sample are

\begin{equation}
\label{eq::bianoiiioii}
    m = 8.54 - (0.59 \times \text{O32}_{\text{B}}),
\end{equation}

\begin{equation}
\label{eq::bianoiiihbeta}
    \text{O3}_{\text{B}} = 43.9836 - 21.6211m + 3.4277m^2 - 0.1747m^3,
\end{equation}

\begin{equation}
\label{eq::bianR23}
    \text{R}_{23} = 138.0430 - 54.8284m + 7.2954m^2 - 0.32293m^3,
\end{equation}

\noindent where $\text{O32}_{\text{B}} = \log_{10}([\text{OIII}]4959,5007/[\text{OII}]3727,3729)$, $\text{O3}_{\text{B}} = \log_{10}([\text{OIII}]4959,5007/\text{H}\beta)$ and $m = 12 + \log_{10}(\text{O/H})$. We determined gas-phase metallicity values following the $\chi^{2}$ minimisation method explained in Section \ref{subsec::maiolinocal}, including the same selection process for when [OII]3727 and/or H$\beta$ are poorly detected (an upper limit on [OII]3727,3729 is used in Equation \ref{eq::bianoiiioii} in the case of both [OII]3727 and H$\beta$ being poorly detected, resulting in an upper limit on metallicity in all four cases where this was used). The solution to Equation \ref{eq::bianoiiioii} was used to discriminate between any multiple solutions. Equation \ref{eq::bianoiiioii} is suitable for $0.3 < \text{O32} < 1.2$ \citep{Bian2018} but here, following the method of \citet{Li2023}, we extrapolated the relationship linearly.
%%%%%%%%%%%%%%%%%%%%%%%%%%%%%%%%%%%%%%%%%%%%%%%%%%%%%%%%%%%%%%%%%%%%%%%%%%%%%%%%%%%%%%%%%%%%%%%%
\section{Results}
\label{sec::results}
%%%%%%%%%%%%%%%%%%%%%%%%%%%%%%%%%%%%%%%%%%%%%%%%%%%%%%%%%%%%%%%%%%%%%%%%%%%%%%%%%%%%%%%%%%%%%%%%
\subsection{Mass-Metallicity Relation}
\label{subsec::mzrelation}

\begin{figure*}
\centering

\begin{subfigure}{0.45\linewidth}
\includegraphics[width=\linewidth, trim=30 0 50 0]{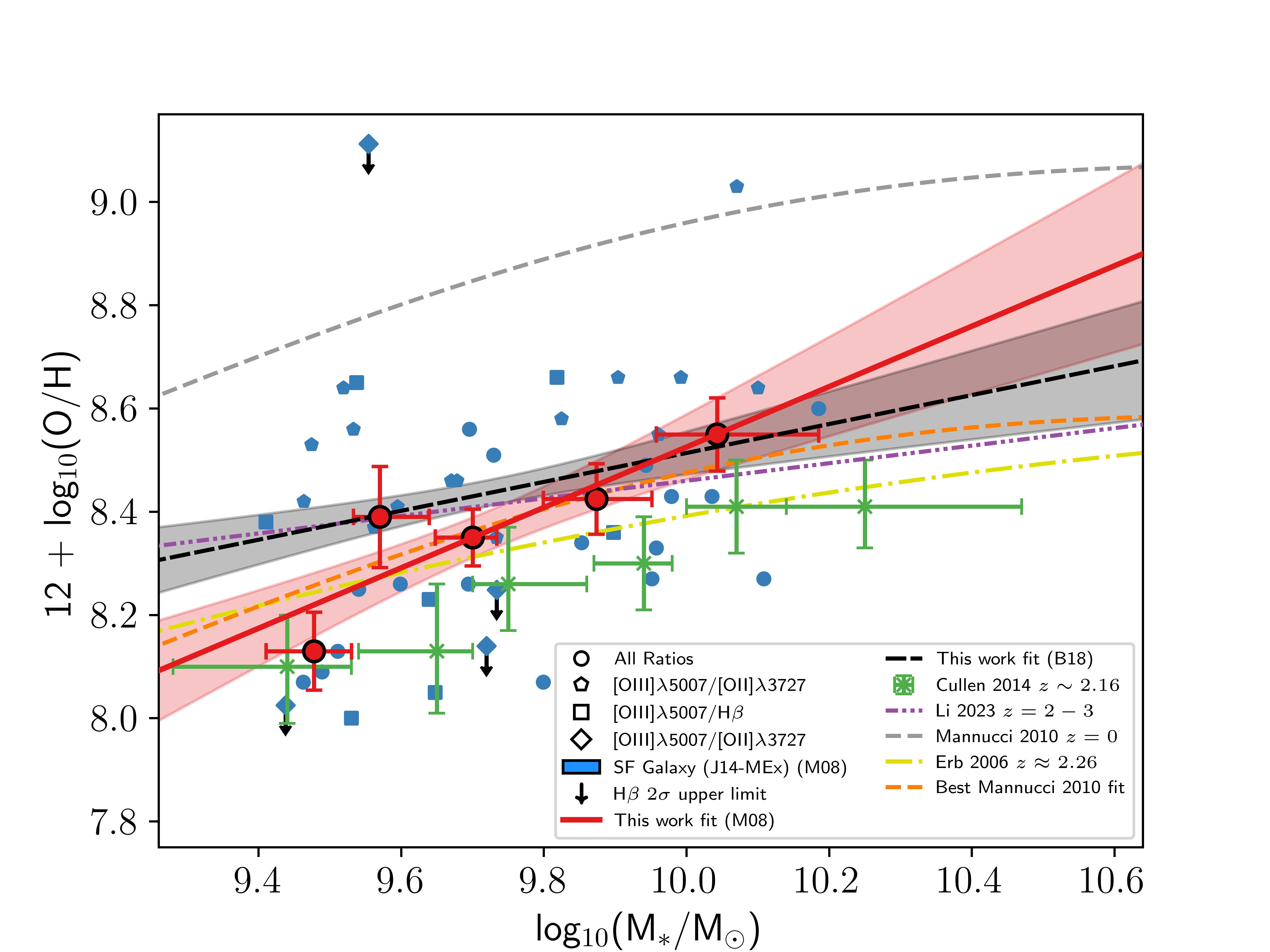}
\caption{ }
\label{subfig::MZRa}
\end{subfigure}
\hfill
\begin{subfigure}{0.45\linewidth}
\includegraphics[width=\linewidth, trim=50 0 30 0]{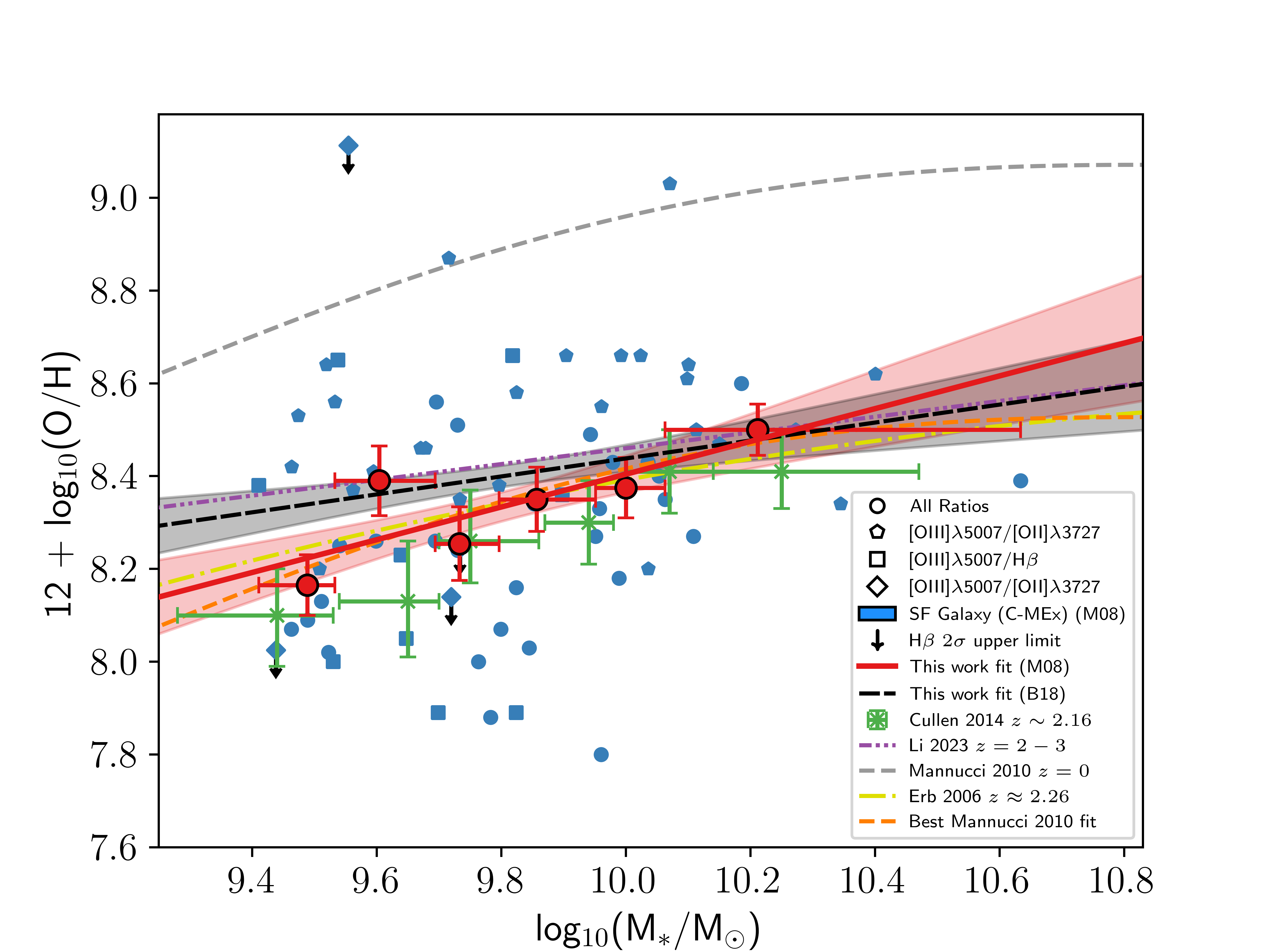}
\caption{ }
\label{subfig::MZRb}
\end{subfigure}

\caption{The mass-metallicity relation (MZR) for our sample. The shape of the blue points (individual SF galaxies calibrated from the \citealp{Maiolino2008} calibrations) are selected based on the strong-line ratio used to determine the galaxy's gas-phase metallicity. Black arrows indicate limits on metallicity. The red circles are the binned averages of the SF galaxies within the completeness limit of our sample (8-9 SF galaxies per bin for the J14-MEx sample, and 12 SF galaxies per bin in the C-MEx sample), with the error on $\log_{10}(\text{M}_{*}/\text{M}_{\odot})$ representing the width of the bin, and error on metallicity being the standard error within that bin. The solid red line shows the fit of the binned averages of our \citet{Maiolino2008} sample (the red shaded region indicates the 1$\sigma$ error of the fit). The green crosses show the \citet{Cullen2014} MZR data. The grey dashed line shows the $z=0$ MZR from \citet{Mannucci2010}, and the orange dashed line shows the offset version of their best fit fitted to our sample. The yellow dash-dot line shows the MZR from \citet{Erb2006}, which is the local \citet{Tremonti2004} MZR shifted by -0.56 dex in metallicity to more closely match their $z\approx2.26$ sample. The purple dash-double dot line shows the linear MZR from \citet{Li2023}. \citet{Erb2006} and \citet{Li2023} used different metallicity calibrations to those of our sample and direct comparisons should be noted with caution. The black dashed line shows the best fit to median bins of our sample of SF galaxies when calibrated using the \citet{Bian2018} metallicity calibrations, which are more applicable to high-$z$ (the black shaded region indicates the 1$\sigma$ scatter of this fit). We do not plot the individual or binned data points of SF galaxies from the \citet{Bian2018} calibrations to avoid confusion for the reader. As with the \citet{Erb2006} and \citet{Li2023} lines, this fit should be noted with caution.}
\label{fig::mzrfull}
\end{figure*}

Figure \ref{fig::mzrfull} shows the MZR of our complete sample compared with other observational works. The left panel shows the MZR for our sample of J14-MEx SF galaxies, and the right panel shows the same but for our sample of C-MEx SF galaxies. The MZR was determined by measuring the best fit of the median stellar mass and gas-phase metallicity values in approximately equal-sized mass bins for both of our samples (8-9 SF galaxies per bin for the J14-MEx sample, and 12 SF galaxies per bin for the C-MEx sample). The error bars on these bins are the width of the bin in mass and the standard error for the median metallicity. These fits are compared with other works. The best fit was found to be linear in the form $a \cdot x + b$, where $x = \log_{10}(\text{M}_{*}/\text{M}_{\odot})$, $a$ is the slope of the fit and $b$ is the intercept. The fit to our sample is shown by the dashed red line in Figure \ref{fig::mzrfull}, with the red shaded region representing the 1$\sigma$ scatter. The properties of these fits to both our J14-MEx and C-MEx samples can be found in Table \ref{tab::mzrthesefits}. From both panels in Figure \ref{fig::mzrfull}, it can be seen that there is a correlation between gas-phase metallicity and stellar mass for the galaxies in our complete sample. The binned medians of our sample in the mass range of \citet{Cullen2014} ($\log_{10}(\text{M}_{*}/\text{M}_{\odot})$ = 9.44 - 10.25) follow the same general trend, but their points are offset by $-0.08\pm0.05$ dex in metallicity in the same mass range for our C-MEx sample ($-0.17\pm0.06$ dex for our J14-MEx sample).

\begin{table}
    
    \footnotesize
    \centering
    \caption{MZR fits for the binned data in Figure \ref{fig::mzrfull} in the form $12 + \log_{10}(\text{O}/\text{H}) = a \cdot \log_{10}(\text{M}_{*}/\si{\solarmass}) + b$. }
    \label{tab::mzrthesefits}
    \begin{tabular}{lcc} % 15 columns, alignment for each{}
        \hline
        \citet{Maiolino2008} Calibration & $a$              & $b$           \\
        \hline
J14-MEx                       & $0.59\pm0.20$  & $2.67\pm1.80$ \\
C-MEx                       & $0.35\pm0.10$  & $4.87\pm1.30$ \\
\hline
        \citet{Bian2018} Calibration & $a$              & $b$           \\
        \hline
J14-MEx                       & $0.28\pm0.10$  & $5.71\pm1.00$ \\
C-MEx                       & $0.19\pm0.10$  & $6.50\pm0.90$ \\
        \hline{}
    \end{tabular}
\vspace{-4mm}
\end{table}

In order to compare our MZR to those of other studies, different forms of the MZR found by these works are fit to our data by allowing their intercept to vary but keeping higher order coefficients fixed. Doing this gives a clearer perspective on how the MZR found in this paper is different to those found in the literature. From the \citet{Mannucci2010} MZR, the best fit of their curve to our sample is offset from the original by $-0.51\pm0.03$ dex in metallicity. This is a similar offset that is suggested by \citet{Erb2006} of $-0.56$ dex from the local \citet{Tremonti2004} MZR. This large offset is likely due to the fact \citet{Mannucci2010} based their MZR on a sample of local galaxies from SDSS which do not reflect the SF population of $z\sim2.2$ galaxies from our sample. It should be noted that comparisons between \citet{Mannucci2010} and \citet{Erb2006} should be made with caution because \citet{Erb2006} use the \citet{Pettini2004} metallicity calibrations whereas \citet{Mannucci2010} use those from \citet{Maiolino2008}.

There is better agreement between the \citet{Li2023} MZR and our data, with just a small average offset from their MRZ to ours of $\approx-0.04$ dex in metallicity for our C-MEx sample ($\approx-0.08$ dex offset for the J14-MEx sample). This consistency should be noted with caution as \citet{Li2023} use metallicity calibrations from \citet{Bian2018} whereas the comparisons above are using results calibrated using \citet{Maiolino2008}. In order to make a more accurate comparison to \citet{Li2023}, we have overlaid the MZR derived from the \citet{Bian2018} for the same samples of galaxies (black dashed line in Figure \ref{fig::mzrfull}). We find the MZR exists at $z\sim2.2$ with these higher redshift calibrations, albeit with a shallower slope for both the J14-MEx and C-MEx sample (see Table \ref{tab::mzrthesefits}). The difference from the \citet{Li2023} MZR is just $\approx-0.03$ dex in metallicity for the C-MEx sample ($\approx0.04$ dex offset for the J14-MEx sample). These offsets are similar and the slopes are much more consistent; the \citet{Li2023} MZR occupies the 1$\sigma$ scatter of our fit for almost the entire mass range of our C-MEx sample.

\begin{figure}
        \centering
        \includegraphics[width=\columnwidth, trim=0 0 0 0, clip=true]{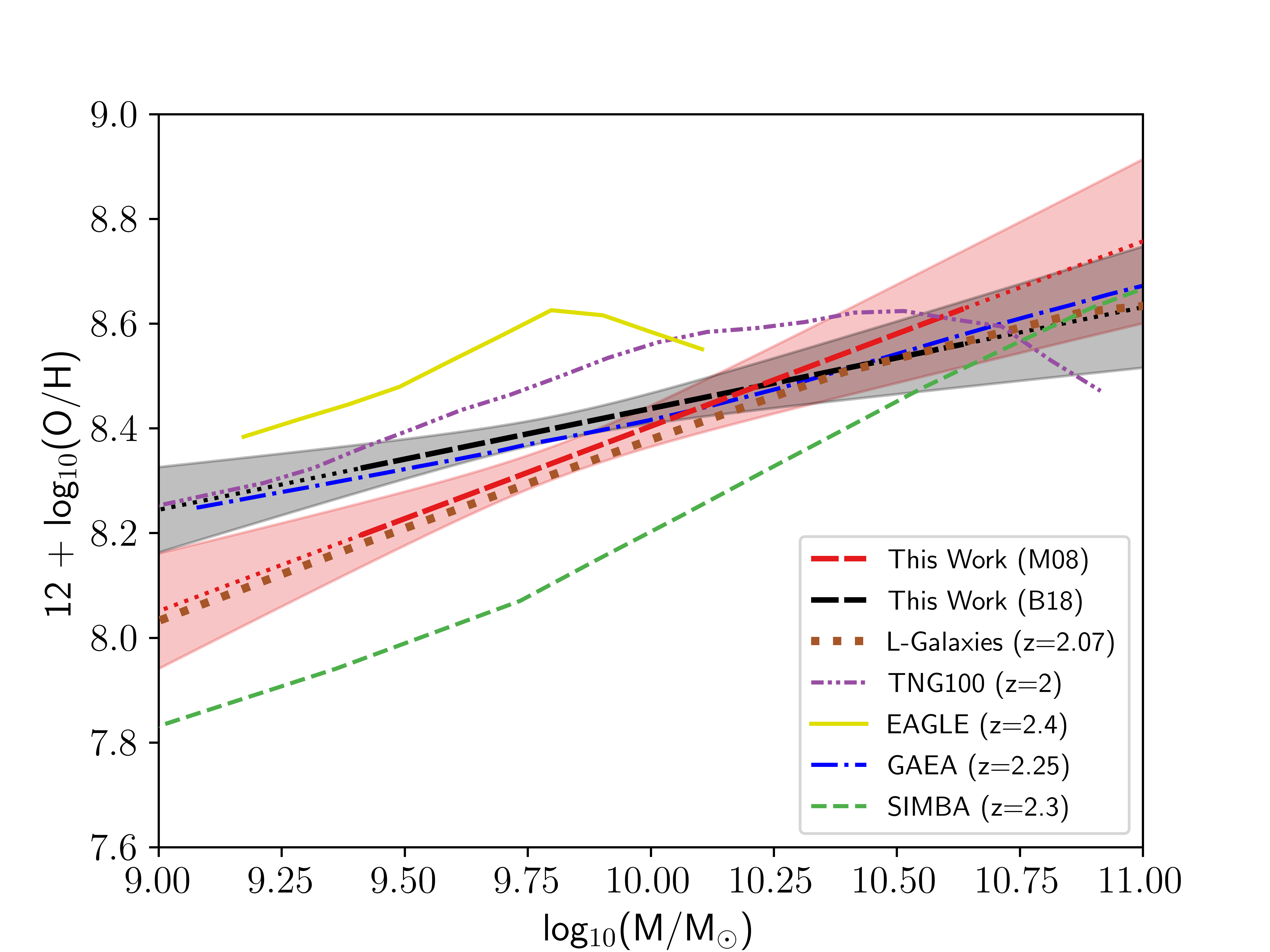}
    \caption[]{The MZR for our sample compared to MZRs generated from simulations of galaxy evolution. The dashed red line shows the fit of the binned averages of our C-MEx sample using metallicity calibrations from \citet{Maiolino2008} (the red shaded region indicates the 1$\sigma$ error of the fit). The dashed black line shows the fit of the binned averages of our C-MEx sample using metallicity calibrations from \citet{Bian2018} (the black shaded region indicates the 1$\sigma$ error of the fit). The brown dotted line indicates the $z=2.07$ binned median MZR relationship of the cosmological semi-analytic galaxy evolution simulation, L-Galaxies from \citet{Yates2023}. The dash-dotted purple line shows the $z=2$ MZR derived from the IllustrisTNG TNG100 cosmological hydrodynamical simulation in \citet{Torrey2019}. The solid yellow line shows the $z=2.4$ MZR determined by \citet{DeRossi2017} using the EAGLE suite of cosmological hydrodynamical simulations. The blue dash-dot shows the shifted $z=2.25$ MZR from analysis by \citet{Fontanot2021} using the GAEA semi-analytical model of galaxy formation. The green dashed line indicates the \citet{Dave2019} $z=2.3$ MZR using the SIMBA cosmological hydrodynamical simulations.}
        \label{fig::simulatoncomparison}
\end{figure}

In Figure \ref{fig::simulatoncomparison}, we compare the MZR of our C-MEx sample using both the \citet{Maiolino2008} (dashed red line) and \citet{Bian2018} (dashed black line) calibrations to results from simulations in the literature at comparable redshifts. These simulations are the $z=2.07$ binned median MZR from L-Galaxies\footnote{https://lgalaxiespublicrelease.github.io} in \citet{Yates2023}, which is a cosmological semi-analytic galaxy evolution simulation \citep{Springel2005a,Henriques2020}; the $z=2$ MZR found by \citet{Torrey2019}, derived from the IllustrisTNG TNG100 cosmological hydrodynamical simulation \citep{Marinacci2018}; the $z=2.4$ MZR determined by \citet{DeRossi2017} using the Evolution and Assembly of GaLaxies and their Environments (EAGLE) suite of cosmological hydrodynamical simulations; the $z=2.25$ MZR from analysis by \citet{Fontanot2021} using the GAEA semi-analytical model of galaxy formation \citep{Hirschmann2016}; and the $z=2.3$ MZR from \citet{Dave2019} found using the SIMBA cosmological hydrodynamical simulations. For the \citet{Fontanot2021} line, we chose to use their MZR that is shifted by $-0.1$ dex in metallicity from their intrinsic model predictions. We selected this instead of the direct model because, as \citet{Fontanot2021} explain, this shift is acceptable given the uncertainty in the normalisation of the MZR from the metallicity indicators, and this offset finds much better agreement with the $z\sim0$ observations they compare to, whilst maintaining agreement with observations at all other redshifts they analyse (see their Section 4.1). From Figure \ref{fig::simulatoncomparison}, the MZR of our C-MEx galaxies for both calibrations generally agree with those measured in simulations at these redshifts, especially at higher masses ($\sim 10^{10.25-10.75}$ \si{\solarmass}), although the EAGLE simulations from \citet{DeRossi2017} only go up to $\sim10^{10.1}$ \si{\solarmass}. The MZR from L-Galaxies \citep{Yates2023}, which is the median binned relationship of their simulated galaxies, agrees remarkably well with our \citet{Maiolino2008} calibrated relationship at low stellar masses ($\sim 10^{9-10}$ \si{\solarmass}), but then strongly agrees with our \citet{Bian2018} calibrated relationship at higher stellar masses ($\sim 10^{10.25-11}$ \si{\solarmass}) following a flattening of their relationship. Additionally, there is strong agreement between the semi-analytic model from \citet{Fontanot2021} and our \citet{Bian2018} calibrated sample for the entire mass range of Figure \ref{fig::simulatoncomparison}, with their MZR not falling outside the 1$\sigma$ scatter of our relationship.

%%%%%%%%%%%%%%%%%%%%%%%%%%%%%%%%%%%%%%%%%%%%%%%%%%%%%%%%%%%%%%%%%%%%%%%%%%%%%%%%%%%%%%%%%%%%%%%%
\subsection{The Fundamental Metallicity Relation}
\label{subsec::fmr}

\citet{Mannucci2010} introduced a way to project the 3D relationship between SFR, stellar mass and gas-phase metallicity onto a 2D plane by combining stellar mass and SFR into a single axis. They describe that this should show a more accurate correlation with gas-phase metallicity because, for a given stellar mass, galaxies with a higher SFR have reduced metallicities and exhibit properties of lower mass galaxies. As a result, while direct relationships between SFR and gas-phase metallicity may show that they are directly correlated properties, they may be masking over more complicated trends when stellar mass is considered as well. This FMR projection is in the form
\begin{equation}
\label{eq::FMReq}
    \mu_{\alpha} = \log_{10}(\text{M}_{*}/\si{\solarmass}) - \alpha\log_{10}(\text{SFR/\si{\solarmass\per\year}}),
\end{equation}

\noindent where $\alpha$ is a free parameter determined by finding the minimum gas-phase metallicity dispersion in this plane. For their data, \citet{Mannucci2010} found $\alpha = 0.32$. For our samples, using the higher redshift metallicity calibrations from \citet{Bian2018}, the scatter around metallicity was minimised for $\alpha = 0.48$ when using the J14-MEx, and $\alpha = 0.65$ for the C-MEx sample. The latter value of 0.65 is consistent with $\alpha$ values obtained for galaxies in this redshift range in the literature (e.g. \citealp{Curti2020a,Sanders2021}; and \citealp{Li2023}, see below). Figure \ref{fig::alphaminimised} shows the metallicity dispersion as a function of $\alpha$ for the C-MEx sample. Despite still indicating the existence of an FMR, an $\alpha$ of $0.48$ for the J14-MEx sample is somewhat at odds with expected values at this redshift range. Randomly sampling the C-MEx sample to match the size of the J14-MEx sample 10,000 times gives an average $\alpha \approx 0.71 \pm 0.19$, which is consistent with the C-MEx value of $\alpha = 0.65$. This suggests that the smaller sample size resulting from the more restrictive J14-MEx prescriptions is not the primary cause of the lower value. The lower $\alpha$ is instead likely due to the J14-MEx samples not probing as wide a mass and SFR range as the C-MEx sample. Figure \ref{fig::fmrthisfluxlimit} shows the FMR using these obtained $\alpha$ values for our [OIII]5007 flux and stellar mass complete sample. The FMR in Figure \ref{subfig::fmrthisfluxlimitb}, which is for our C-MEx sample, is constructed in the form of
\begin{equation}
\label{eq::fluxlimitfmrcoil}
    12 + \log_{10}(\text{O}/\text{H}) = (4.8\pm0.9) + (0.4\pm 0.1)\mu_{0.65}.
\end{equation}

This linear form is fit to equally-sized bins of our complete sample showing the median metallicity of the galaxies in each bin. The significance of the slope of this relation is $\sim4\sigma$, calculated by dividing the slope of the fit by its uncertainty. The shaded region around the red dashed line indicates the 1$\sigma$ uncertainty about the fit.

\begin{figure}
        \centering
        \includegraphics[width=\columnwidth, trim=0 0 0 0, clip=true]{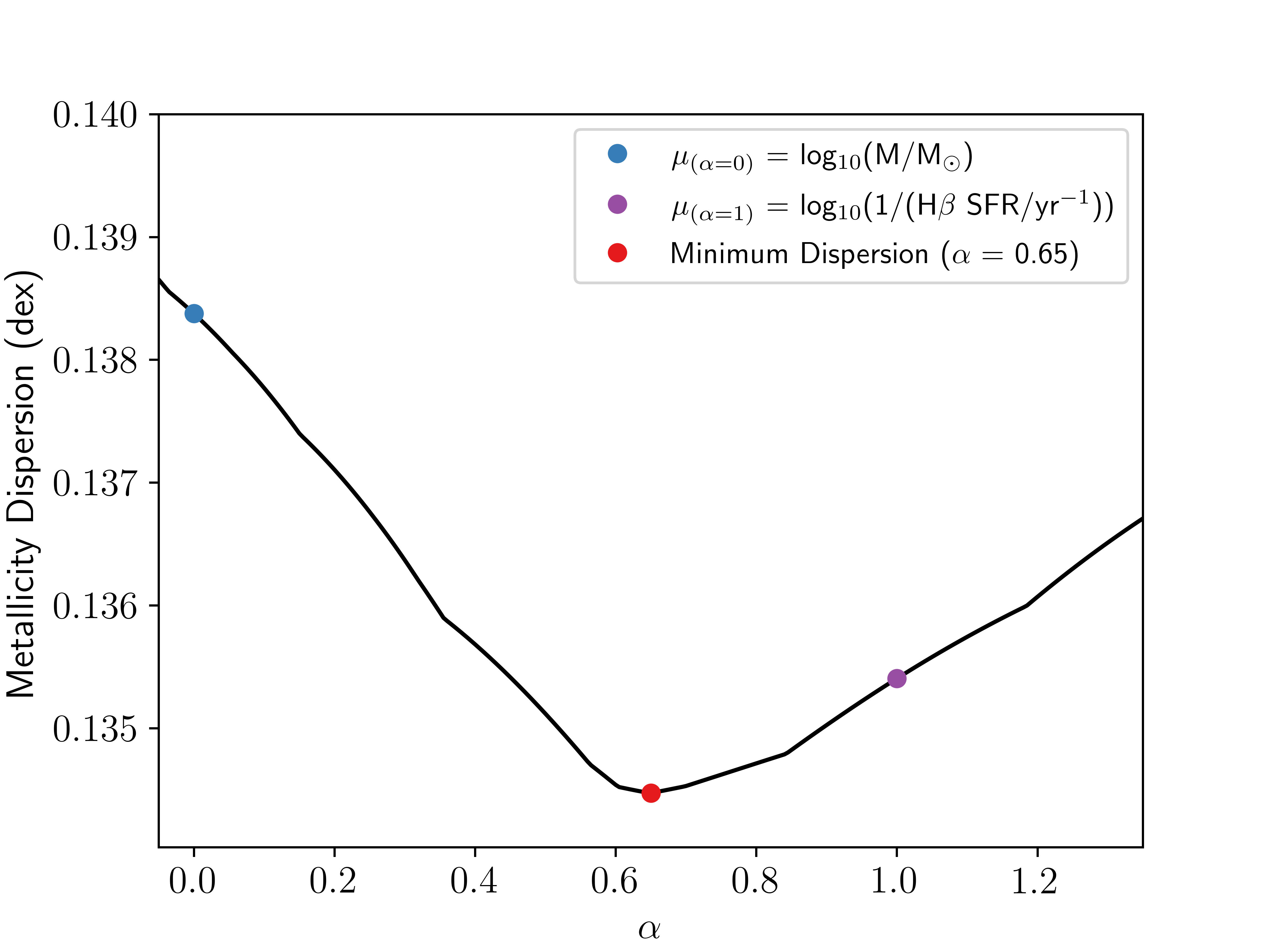}
    \caption[]{Metallicity dispersion of SF galaxies defined using the C-MEx as a function of $\alpha$ (defined in Equation \ref{eq::FMReq}). Shown are $\alpha$ values corresponding to the minimum dispersion about metallicity ($\alpha = 0.69$, red point), $\alpha = 0$ ($\mu_{0} = \log_{10}(\text{M}_{*}/\si{\solarmass})$, blue point) and $\alpha = 1$ ($\mu_{1} = \log_{10}(\frac{1}{\text{H}\beta\text{ sSFR}/\text{yr}^{-1}})$, purple point).}
        \label{fig::alphaminimised}
\end{figure}{}

Comparisons with the FMRs of other works can be done by using their values of $\alpha$, making sure to maintain the same calibrations for metallicity calculations. Figure \ref{fig::mannuccifmr} shows our samples on the FMR plane of \citet{Mannucci2010}, with the purple fit showing their general form of the FMR for galaxies of any stellar mass, SFR and redshift up to $z\approx2.5$. We used the metallicity calibrations from \citet{Maiolino2008} for this comparison. The \citet{Mannucci2010} FMR is in the form
\begin{equation}
\label{eq::manngeneralform}
       12 + \log_{10}(\text{O}/\text{H}) =
    \begin{cases}
        8.90 + 0.47(\mu_{0.32} -10) & \text{if $\mu_{0.32} < 10.2$} \\
        9.07 & \text{if $\mu_{0.32} > 10.5$}.
    \end{cases}
\end{equation}

All the galaxies in our sample have $\mu_{0.32} < 10.2$, so for simplicity we will represent this as $12 + \log_{10}(\text{O}/\text{H}) = 4.20 + 0.47\mu_{0.32}$. From Figure \ref{subfig::mannucifmrcoil}, using $\alpha = 0.32$ and $x = \mu_{0.32}$, the FMR for our sample of C-MEx SF galaxies is found to be
\begin{equation}
\label{eq::mannuccifmrthiswork}
    12 + \log_{10}(\text{O}/\text{H}) = (3.92\pm1.0) + (0.47\pm0.10)\mu_{0.32}.
\end{equation}

The slope is in very good agreement with the general linear form found by \citet{Mannucci2010}, but is offset by $\sim 0.28\pm0.04$ dex in metallicity in the $\mu_{0.32}$ range of the sample. The error on this offset is the median 1$\sigma$ error about our fit in this range. This offset is in good agreement with \citet{Cullen2014} who found that the FMR of their sample of $z_{\text{median}} \sim 2.16$ SF galaxies, also observed with \emph{HST} grism, was offset by an average of $\sim0.3$ dex. They determined that this discrepancy is due to the selection of metallicity indicator because \citet{Maiolino2008} relies on local SF galaxies to determine their metallicity calibrations, which may not be applicable at high-$z$, and our result adds evidence to this suggestion.

\begin{figure*}
\centering

\begin{subfigure}{0.45\linewidth}
\includegraphics[width=\linewidth, trim=0 0 50 0]{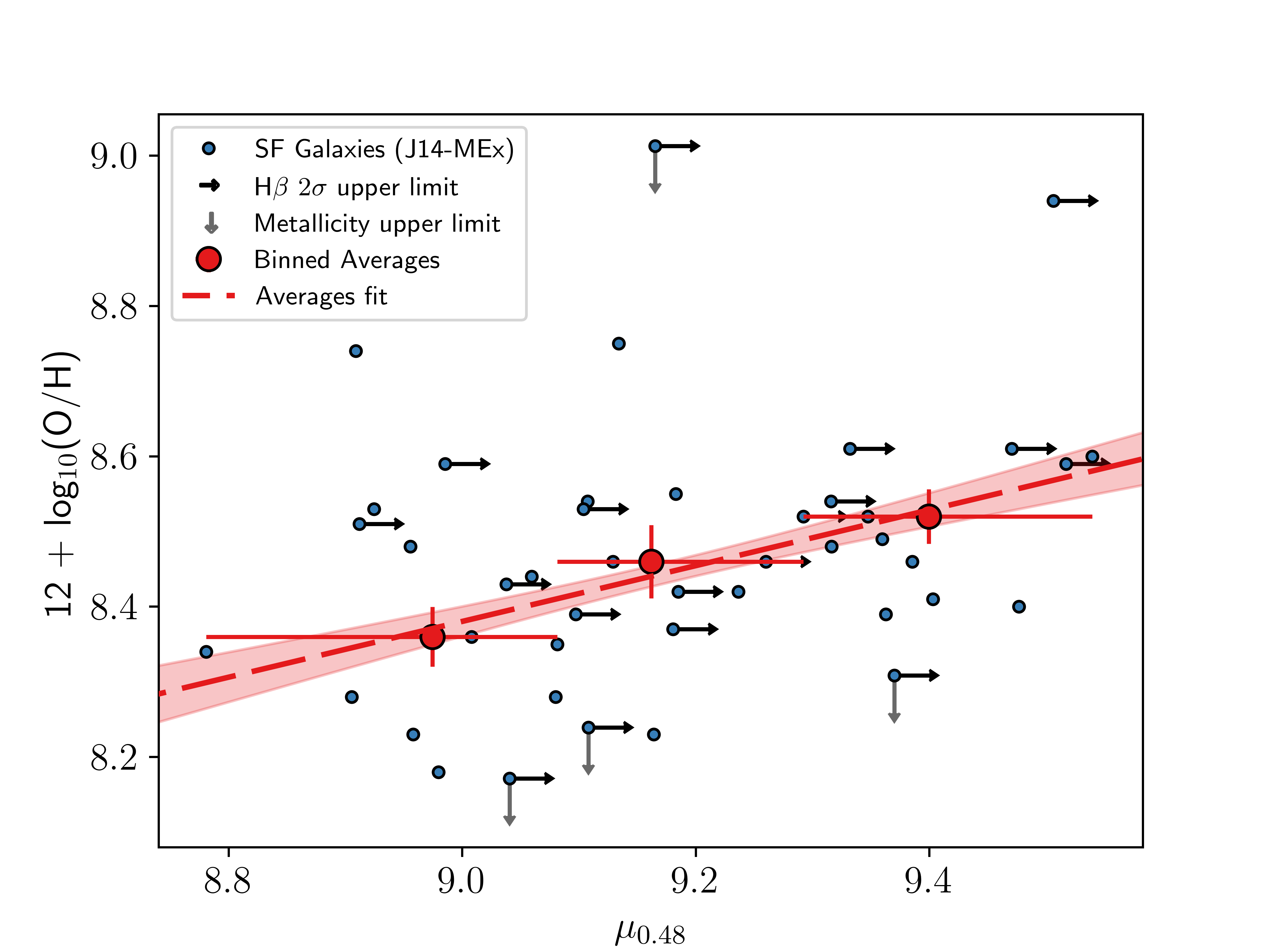}
\caption{ }
\label{subfig::fmrthisfluxlimita}
\end{subfigure}
\hfill
\begin{subfigure}{0.45\linewidth}
\includegraphics[width=\linewidth, trim=50 0 0 0]{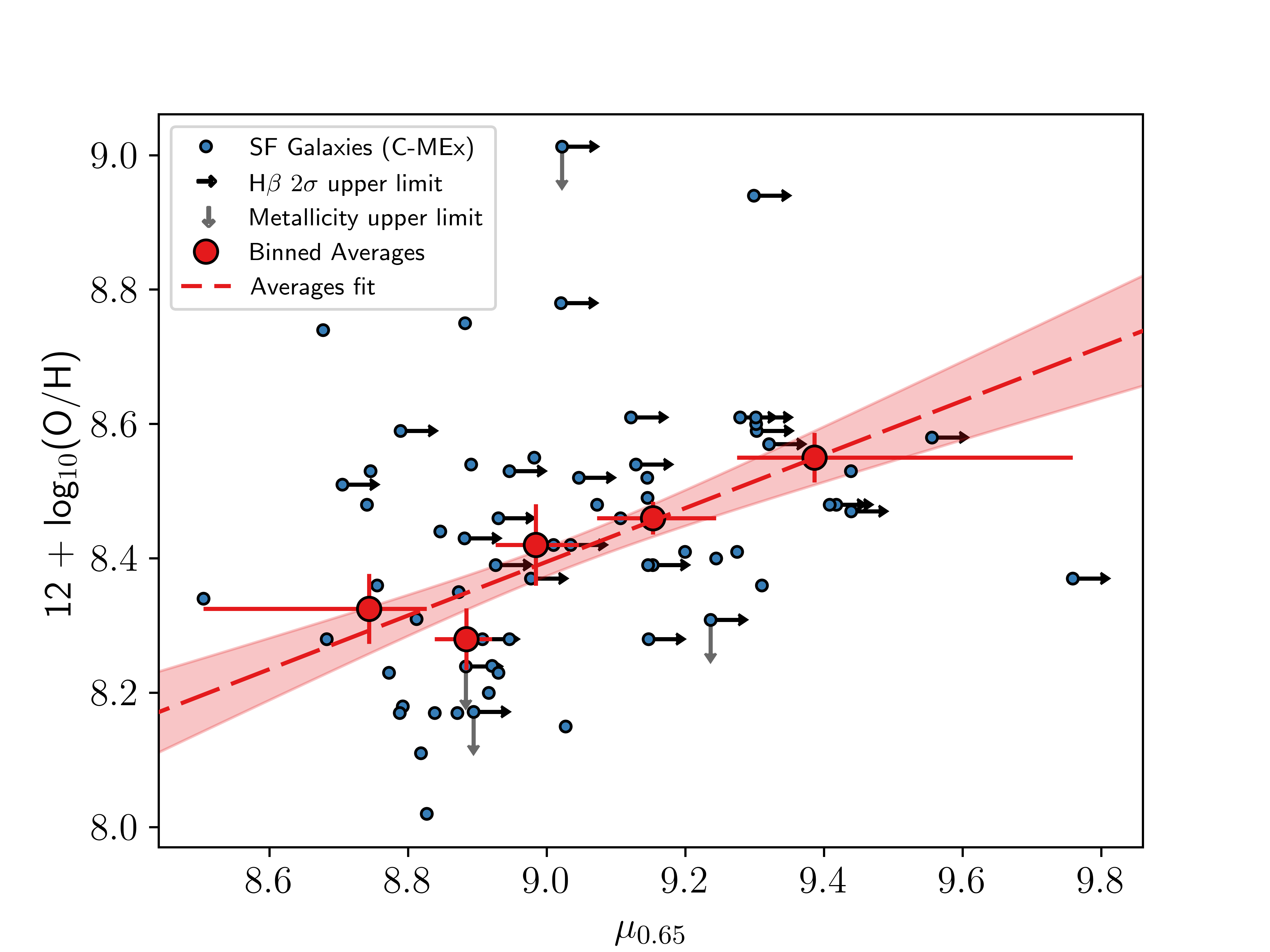}
\caption{ }
\label{subfig::fmrthisfluxlimitb}
\end{subfigure}

\caption{The 2D projection of the FMR using the form $\mu_{\alpha} = \log_{10}(\text{M}) - \alpha\log_{10}(\text{SFR})$ as introduced in \citet{Mannucci2010}. The $\alpha$ values in both panels are those calculated when the scatter about gas-phase metallicity is minimised. \emph{Left} - The 2D-projected FMR using SF galaxies as defined by the shifted J14-MEx curve, where $\alpha = 0.48$. \emph{Right} - The 2D-projected FMR using SF galaxies as defined by the C-MEx, where $\alpha = 0.65$. In both panels, the blue points show the individual SF galaxies in our complete sample. The red points show the median values for our complete sample in approximately-equal sized $\mu_{\alpha}$ bins, with the red dashed line showing the best fit to these points (the red shaded regions indicate the 1$\sigma$ scatter about the fit).}
\label{fig::fmrthisfluxlimit}
\end{figure*}

\begin{figure*}
\centering

\begin{subfigure}{0.45\linewidth}
\includegraphics[width=\linewidth, trim=0 0 50 0]{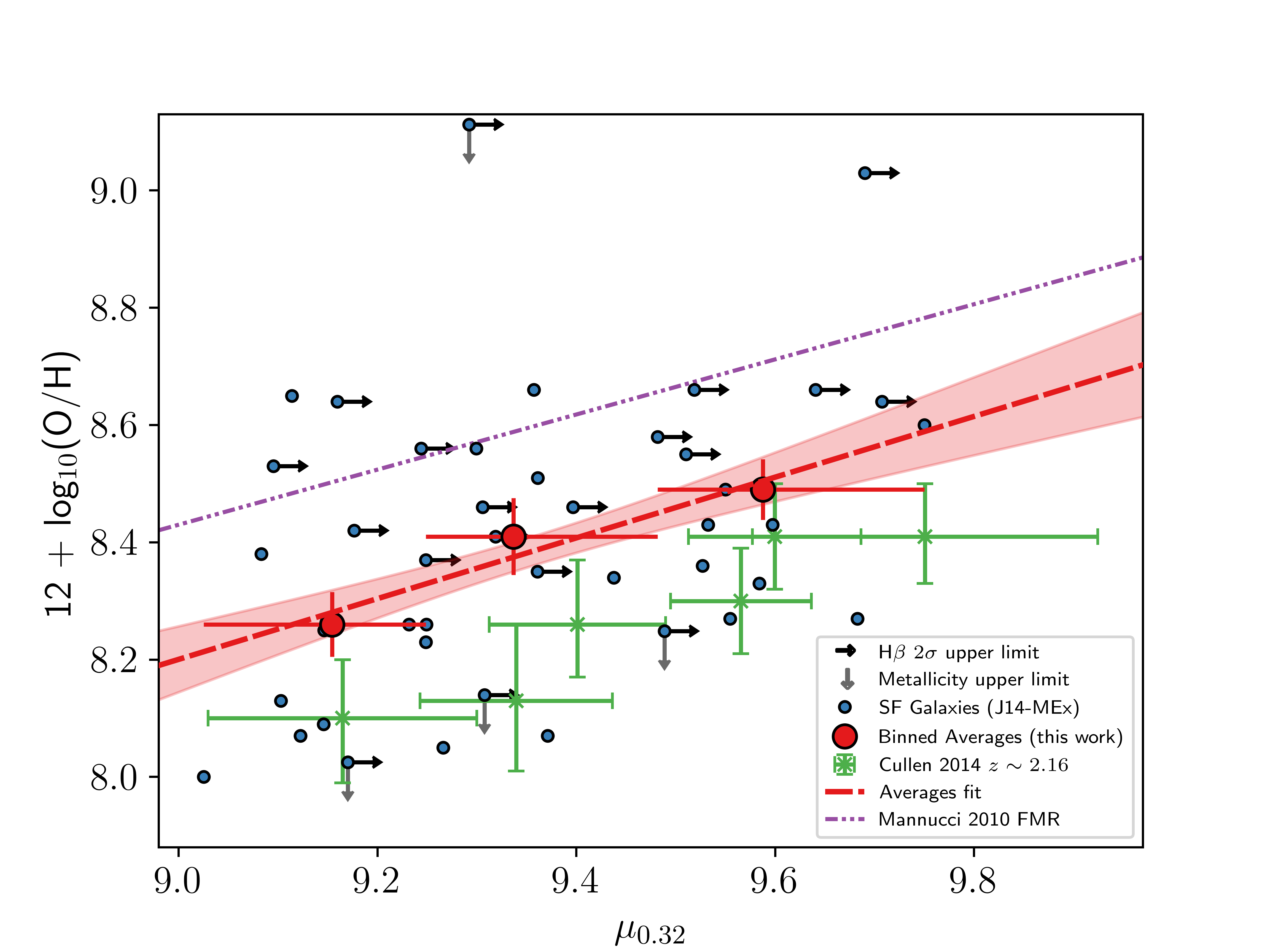}
\caption{ }
\label{subfig::mannuccifmrjun}
\end{subfigure}
\hfill
\begin{subfigure}{0.45\linewidth}
\includegraphics[width=\linewidth, trim=50 0 0 0]{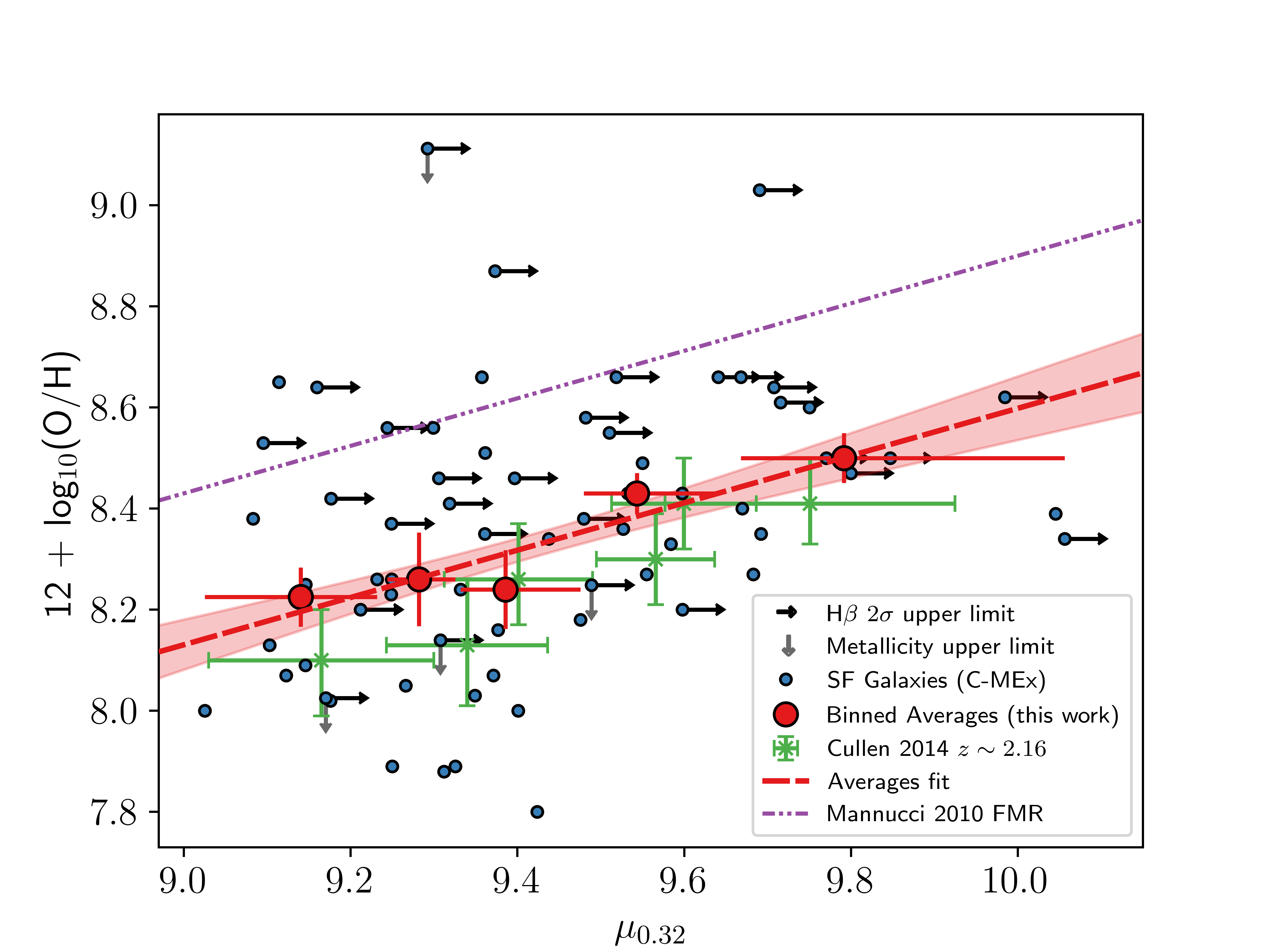}
\caption{ }
\label{subfig::mannucifmrcoil}
\end{subfigure}

\caption{The 2D projection of the FMR using the form $\mu_{0.32} = \log_{10}(\text{M}) - 0.32\log_{10}(\text{SFR}) - 10$ as used by \citet{Mannucci2010}. $\alpha = 0.32$ was found to minimise the scatter on gas-phase metallicity of local SDSS galaxies by \citet{Mannucci2010}. \emph{Left} - The 2D-projected FMR using SF galaxies from our sample as defined by the shifted J14-MEx. \emph{Right} - The 2D-projected FMR using SF galaxies from our sample as defined by the C-MEx. In both panels, the purple dash-double-dot line shows the exact form found by \citet{Mannucci2010} for galaxies of all stellar mass, SFR and at any redshift when $\mu_{0.32} < 10.2$. The blue points show the individual SF galaxies within our complete sample. The red points show the median values for SF galaxies in our complete sample in equal-sized $\mu_{0.32} - 10$ bins, with the red dashed line showing the best fit to these points (the red shaded region indicates the 1$\sigma$ scatter about the fit). The green crosses show the data from \citet{Cullen2014} in this plane.}
\label{fig::mannuccifmr}
\end{figure*}

\begin{figure*}
\centering

\begin{subfigure}{0.45\linewidth}
\includegraphics[width=\linewidth, trim=0 0 50 0]{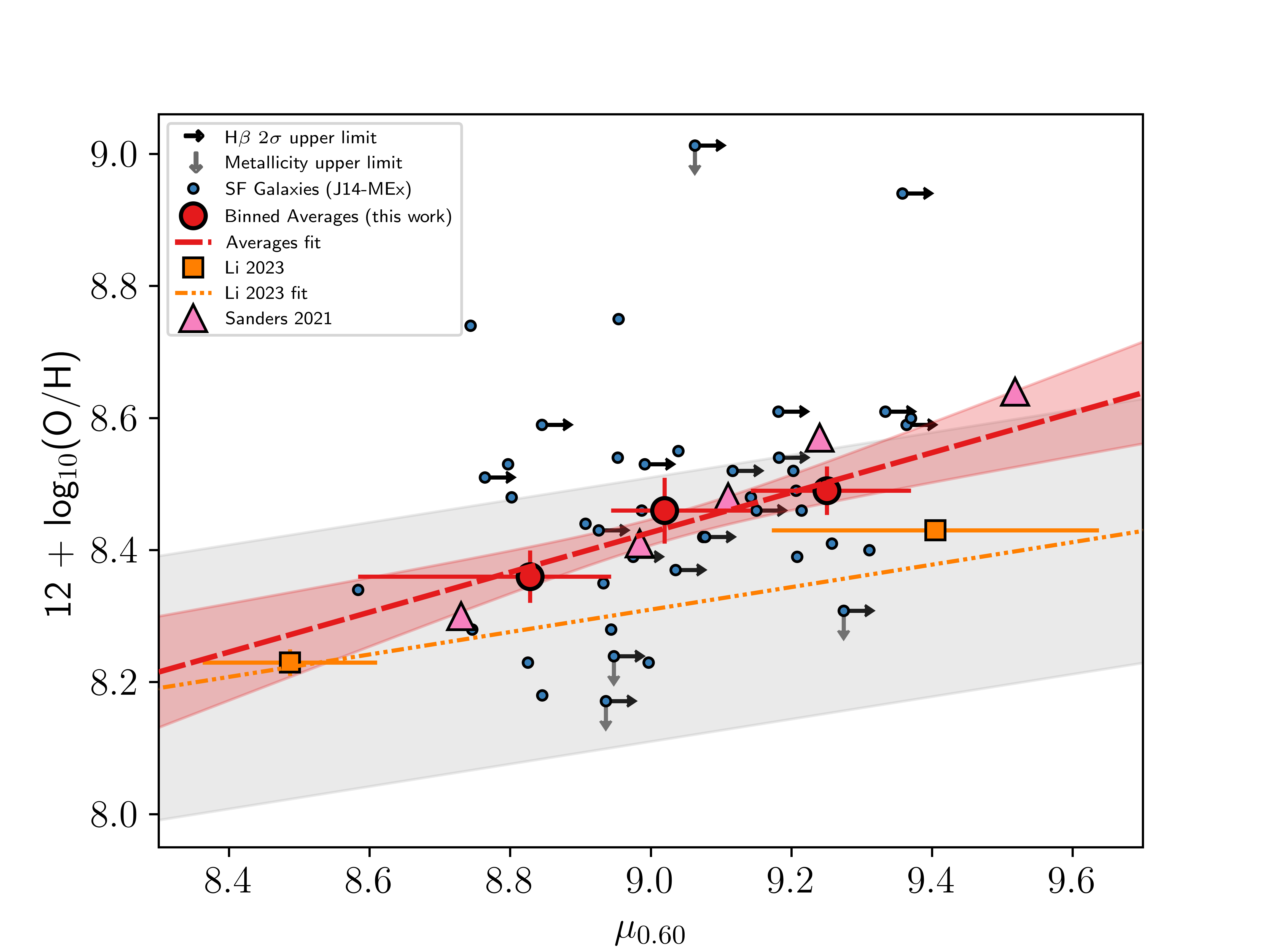}
\caption{}
\label{subfig::lifmrjun}
\end{subfigure}
\hfill
\begin{subfigure}{0.45\linewidth}
\includegraphics[width=\linewidth, trim=50 0 0 0]{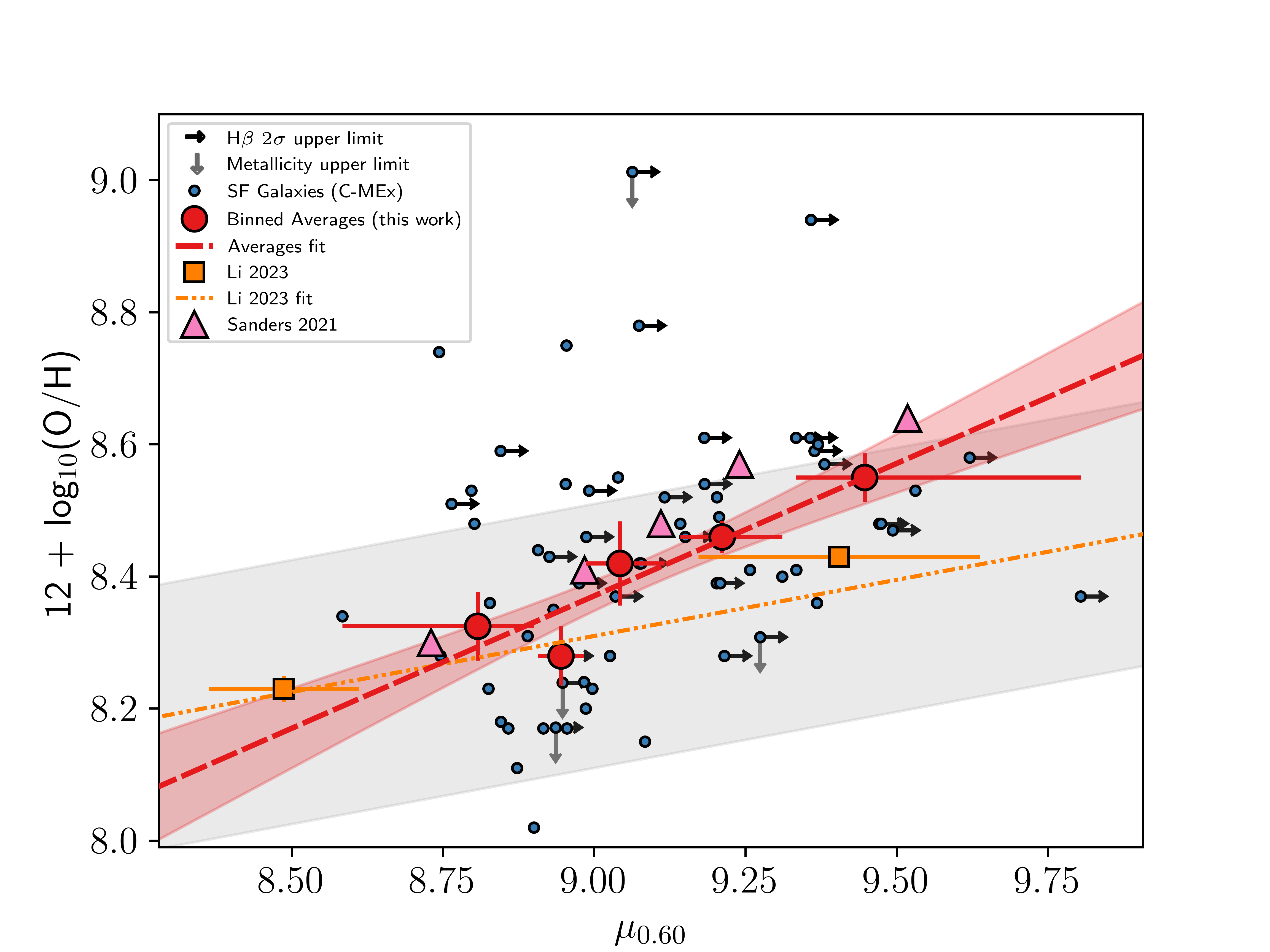}
\caption{}
\label{subfig::lifmrcoil}
\end{subfigure}

\caption{The 2D projection of the FMR using the form $\mu_{0.60} = \log_{10}(\text{M}) - 0.60\log_{10}(\text{SFR})$ as used by \citet{Li2023}. $\alpha = 0.60$ was found to minimise the scatter on gas-phase metallicity of the $z=2-3$ galaxies in \citet{Li2023}. \emph{Left} - The 2D-projected FMR using SF galaxies from our sample as defined by the shifted J14-MEx. \emph{Right} - The 2D-projected FMR using SF galaxies from our sample as defined by the C-MEx. In both panels, the orange squares show the median values from \citet{Li2023} as defined in their paper, with the orange dash-double-dot line showing their fit (grey shaded region indicates the 1$\sigma$ scatter about their fit). The pink triangles show the median values of data from \citet{Sanders2021} in this plane. The blue points show the individual SF galaxies in our complete sample. The red points show the median values in equal-sized $\mu_{0.60}$ bins for our complete sample, with the red dashed line showing the best fit to these points (the red shaded region indicates the 1$\sigma$ scatter about our fit).}
\label{fig::lifmr}
\end{figure*}

Figure \ref{fig::lifmr} shows the FMR found by \citet{Li2023} for $z = 2 - 3$ dwarf galaxies (brown dash-double dot line). Using their value of $\alpha = 0.6$ for the galaxies in our C-MEx sample, and using the \cite{Bian2018} metallicity calibrations as they do, yields an FMR in the form
\begin{equation}
\label{eq::lifmrthiswork}
    12 + \log_{10}(\text{O}/\text{H}) = (4.76\pm0.9) + (0.40 \pm 0.1)\mu_{0.60}.
\end{equation}

This form is a much steeper slope than the trend found by \citet{Li2023} of ($0.17\pm0.02$)$\mu_{0.60}$. This could be caused by the contributions of their lower mass bins ($\approx 10^{6.5} - 10^{7.6}$ \si{\solarmass}) which extend beyond the lower limit of masses in this work. Instead, a more appropriate comparison would be to the $z\sim2.3$ \citet{Sanders2021} sample (pink triangles) in Figure \ref{fig::lifmr} as they all fall within the $\mu_{0.60}$ range of this work (red points) and were also used for comparison in \citet{Li2023}. \citet{Sanders2021} investigated the redshift evolution of the MZR by analysing samples of galaxies at $z\sim2.3$ and $z\sim3.3$ from the MOSFIRE Deep Evolution Field (MOSDEF) Survey (see \citealp{Kriek2015}). Comparisons are made to their $z\sim2.3$ sample as it is a similar redshift to the galaxies in our complete sample and they use the metallicity calibrations from \citet{Bian2018}. A fit to their sample in the same form as the FMR from \citet{Li2023} yields
\begin{equation}
\label{eq::sanderslifmr}
        12 + \log_{10}(\text{O}/\text{H}) = (4.39\pm0.4) + (0.45 \pm 0.04)\mu_{0.60},
\end{equation}
which is in good agreement with our work. This is more evidence that the FMR exists for star-forming galaxies at this redshift.

\citet{Li2023} discuss the possible reasons why their slope of the MZR is shallower than that of \citet{Sanders2021}. They suggest that the evolution of the MZR slope may be determined by different feedback mechanisms and wind models, both of which regulate the fraction of gas ejected from a galaxy \citep{Wang2022} and determine the MZR in individual galaxies. If low-mass galaxies are dominated by different feedback mechanisms compared to high-mass galaxies, then an evolution in the MZR will be visible. They comment that the results of their study are consistent with that of a momentum-driven wind model (see \citealp{Finlator2008,Guo2016}) but at odds with other studies that analyse the MZR at low mass (e.g. \citealp{Torrey2019}). Given our FMR results agree with \citet{Sanders2021} for an overlapping $\mu_{0.60}$ range, but see a steepening of the slope compared to the lower range of \citet{Li2023}, it seems to suggest that there is a turnover in these relationships from low to high mass at $z\sim2.2$. However, while the reasons above for a steepening of the slope in the MZR could apply to a change in slope in the FMR toward higher $\mu_{\alpha}$ values, \citet{Li2023} highlight the need for further analysis of the MZR in dwarf galaxies at these redshifts before making any firm conclusions. Additionally, the fact our \citet{Bian2018} MZR slopes agree with \citet{Li2023} indicates that the above explanations regarding different wind models should not be confidently applied to the FMR.

\section{Discussion \& Conclusions}
\label{sec:conc}

This work analyses the relationship between gas-phase metallicity, stellar mass and SFR in $z=1.99-2.32$ SF galaxies from QSAGE. We used the commonly used [OII]3727,3729, [OIII]4958,5007 and H$\beta$ strong-emission lines in order to calibrate metallicities \citep{Maiolino2008,Bian2018} as they lie in the wavelength range of \emph{HST} grism spectra at these redshifts. Since many strong-emission lines that are used to generate BPT diagrams \citep{Baldwin1981} are unavailable in the wavelength range of our grism spectra, MEx diagrams from \citet{Juneau2014} and \citet{Coil2015} were used to differentiate SF galaxies from AGN. These MEx diagrams use the O3 strong-emission line ratio. As a result of the nature of the observations, [OIII]5007 flux and stellar mass limits were applied in order to negate selection effects that arise from incomplete bins in these two properties. SF galaxies were binned in equal-sized mass bins and an MZR was constructed and compared to previous works in the literature, including MZRs constructed using $z\sim0$ galaxies. In the analysis of the FMR, SF galaxies are binned in equal sized $\mu_{\alpha}$ bins (see Equation \ref{eq::FMReq}), where $\alpha$ varies depending the scatter in gas-phase metallicity. The conclusions of this work are as follows:

\begin{enumerate}[label=\roman*), leftmargin=*]

    \item Using the metallicity calibrations from \citet{Maiolino2008} - making use of four different strong-emission line ratios in our available wavelength range - an MZR can be seen for SF galaxies at the redshift range of our sample. MZRs were built using both of the two MEx diagnostics from \citet{Juneau2014} and \citet{Coil2015} using measured values and upper limits for H$\beta$ flux in the MEx plane. In both of the SF galaxy samples (Figure \ref{fig::mzrfull}), the MZR constructed from this work is consistent with MZRs in the literature for similar redshift ranges \citep{Erb2006,Cullen2014}.

    \item We compared the MZR of our C-MEx sample to a variety of cosmological hydrodynamical simulations and semi-analytic models at comparable redshifts ($z=2-2.4)$, using both the \citet{Maiolino2008} and \citet{Bian2018} metallciity calibrations. We find that, in general, these simulations are consistent with our MZR, particulary at higher stellar masses of $\approx 10^{10.25-10.75}$ \si{\solarmass}.
    
    \item The FMR was investigated using a 2D projection that combines stellar mass and SFR (see \citealp{Mannucci2010}). Analysis using $\alpha=0.65$, a value which minimises the dispersion about gas-phase metallicity for C-MEx galaxies, yields a $\sim4\sigma$ slope for the best fit in this FMR plane. This value of $\alpha$ agrees broadly with recent values in the literature at this redshift. Using the J14-MEx, the value for $\alpha$ which minimises dispersion around gas-phase metallicity comes to $\alpha=0.48$. Despite not agreeing with $\alpha$ values at this redshift in the literature, this fit is still consistent with an FMR being present in our sample. The lower value of $\alpha$ could be due to the J14-MEx not probing as wide a stellar mass and SFR range.

    \item In the \citet{Mannucci2010} FMR plane, where $\alpha=0.32$, our results show that the FMR exists for both J14-MEx and C-MEx diagnostics (Figure \ref{fig::mannuccifmr}). Using the C-MEx, the slope for our sample came to $0.47\pm0.08$, a value that is in excellent agreement for the SDSS sample in \citet{Mannucci2010}. The FMR is offset by $0.28\pm0.04$ dex in metallicity, which is consistent with the offset found by \citet{Cullen2014}. According to \citet{Mannucci2010}, this FMR should be consistent for all masses, SFRs, and redshifts so this offset could be a result of the choice of metallicity calibration at this redshift, a conclusion \citet{Cullen2014} (who similarly used the \citealp{Maiolino2008} calibrations) came to.

    \item Using $\alpha=0.6$ as was used by \citet{Li2023}, the FMR is again found for both J14-MEx and C-MEx SF galaxies. For C-MEx galaxies, the slope of the FMR is found to be $0.40\pm0.1$. This is similar to the slope in the \citet{Mannucci2010} plane but much steeper than the slope found by \citet{Li2023}. This is likely due to the fact their stellar mass range extends down by an additional $\sim3$ dex compared to our sample. The slope is in very good agreement with a fitted slope to the \citet{Sanders2021} data in the same plane ($0.45\pm0.04$), who had a $\mu_{0.60}$ range that overlapped completely with that of this work. The FMR being visible in all three planes analysed in this work strongly suggests that it exists in SF galaxies at this redshift.

\end{enumerate}

Regarding the FMR, we believe the negative correlation with SFR at fixed mass is due to the accretion of metal-poor gas fuelling SFR at cosmic noon (e.g. \citealp{Keres2005,Dekel2009}). There is mounting evidence in the literature that metallicity gradients of higher sSFR galaxies are flat or even positive (i.e. lower metallicity in the central region and increasing with radius; \citealp{Stott2014}. See also \citealp{Wang2017,Gillman2020,Gillman2022}). It is believed these positive metallicity gradients are caused by this metal-poor gas accretion being focussed on the centre of the galaxy \citep{Sharda2021} which can be triggered by either efficient accretion \citep{Stott2014} or merger events \citep{Rupke2010}. Accretion of low-metallicity gas like this results in a dilution of the chemical-abundance. Recently, \citet{Heintz2023} used public \emph{JWST} Near-Infrared Spectrograph datasets to analyse $z>7$ galaxies, and suggested that the drop in gas-phase metallicity they see in these very high redshift galaxies is a result of dilution caused by accretion. Funnelled accretion then drives high sSFR in the centre of galaxies and is thought to be the cause of a relationship between sSFR and metallicity gradient \citep{Wuyts2016,Curti2020b}. In this context, metal-poor gas being efficiently accreted into the central cores of the galaxies greatly enhances the SFR whilst simultaneously reducing the overall average gas-phase metallicity of a galaxy on short timescales, which drives the negative correlation between SFR and gas-phase metallicity for fixed stellar mass we see in the FMR (see also \citealp{Troncoso2014,Kashino2017,Wang2019,Simons2021}). Results from simulations of galaxy evolution lend support to this. From analysis of galaxies in the EAGLE simulations, \citet{DeRossi2017} found that for lower mass galaxies (M$_{*} < 10^{10.3}$ \si{\solarmass}), their results indicate higher fractions of metal-poor gas drive higher sSFRs and reduce gas-phase metallicity values, and that this is regulated by metal-poor inflows, with a particular focus on satellite galaxies. It is worth noting that for their higher mass systems (M$_{*}\gtrsim 10^{10.3}$ \si{\solarmass}), the impact of AGN feedback becomes much more significant to the point where it causes an inversion in the MZR plane for fixed stellar mass; metallicity starts increasing with sSFR for fixed stellar masses at $\gtrsim10.3$ \si{\solarmass}, but at lower masses it is an anti-correlation at fixed stellar mass (they also found a similar inversion at fixed stellar mass for the gas fraction of the SF component of gas). \citet{Torrey2019} found that accretion plays a significant role shaping the FMR in SF galaxies in TNG100, and that, additionally, the MZR is a consequence of the accretion (and enrichment) history of galaxies.

\section*{Acknowledgements}

{The authors gratefully acknowledge the significant contribution to this project made by the late Richard G. Bower. His scientific expertise and enthusiasm are sorely missed.

The authors thank the anonymous referee for their helpful comments that have increased the clarity of this work.

HMOS gratefully acknowledges support from an STFC PhD studentship and the Faculty of Science and Technology at Lancaster University.

MF and RD would like to acknowledge funding from the European Research Council (ERC) under the European Union's Horizon 2020 research and innovation programme (grant agreement No 757535).

For the purpose of open access, the authors have applied a creative commons attribution (CC BY) licence to any author accepted manuscript version arising.}

\section*{Data availability}

The data underlying this article are from \emph{HST} Cycle 24 proposal 14594: ‘QSAGE: QSO Sightline And Galaxy Evolution’ and are publicly available from the Mikulski Archive for Space Telescopes (MAST, \url{https://archive.stsci.edu/hst/}).

%%%%%%%%%%%%%%%%%%%%%%%%%%%%%%%%%%%%%%%%%%%%%%%%%%



\bibliographystyle{mnras}
\bibliography{HMO_MZ_FMR_arxiv.bib} % if your bibtex file is called example.bib


%%%%%%%%%%%%%%%%%%%%%%%%%%%%%%%%%%%%%%%%%%%%%%%%%%

%%%%%%%%%%%%%%%%% APPENDICES %%%%%%%%%%%%%%%%%%%%%


%%%%%%%%%%%%%%%%%%%%%%%%%%%%%%%%%%%%%%%%%%%%%%%%%%

% Don't change these lines
\bsp    % typesetting comment
\label{lastpage}
\end{document}